\documentclass[preprint,12pt]{elsarticle}

\graphicspath{{figures/}}
\usepackage{amstext}

\usepackage{hyperref}

\usepackage{amssymb}

\usepackage{subfigure}

\usepackage{lineno}

\usepackage{xcolor}
\usepackage{xifthen}
\newcommand{\cmnt}[2][]{
    {\ifthenelse
            {\isempty{#1}}
            {\color{red}{#2}}
            {\color{red}{#2}\footnote{\color{red}{#1}}}
        }
}
\newcommand{\cmntd}[2][]{
    {\ifthenelse
            {\isempty{#1}}
            {\color{black}{#2}}
            {\color{black}{#2}$\!\!\!\!\!$}
        }
}

\definecolor{color}{rgb}{1,0,1}

\newcommand{\cfnd}[1]{{}}

\newcommand{\editsf}[1]{{\color{black}{#1}}}

\usepackage{scalerel}
\newcommand{\chr}{{\Theta_{\scaleto{C/H}{4pt}}}}

\usepackage{pifont}
\usepackage{textcomp}
\newcommand{\Nc}{{N_{C}}}

\newcommand{\Nfive}{{N_5}}
\newcommand{\Nsix}{{N_6}}
\newcommand{\Nseven}{{N_7}}
\newcommand{\Nring}{{N_{\scriptscriptstyle\bigcirc}}}
\newcommand{\Ncring}{{N_{\copyright}}}
\newcommand{\Ncnonring}{{N_{\mbox{\textcent}}}}

\usepackage[normalem]{ulem}
\usepackage{nicefrac}

\usepackage{eqnarray}
\usepackage{multirow}
\usepackage{titlesec}
\usepackage{caption}

\usepackage{newfloat}
\DeclareFloatingEnvironment[fileext=frm,placement={!ht},name=Correlation Set]{Cor}
\begin{document}

\begin{frontmatter}

    \title{Physical, Chemical and Morphological Evolution of Incipient Soot Obtained from Molecular Dynamics Simulation of Acetylene Pyrolysis}

    \author[inst1]{Khaled Mosharraf Mukut}

    \affiliation[inst1]{organization={Department of Mechanical Engineering},
        addressline={ Marquette University},
        city={Milwaukee},
        state={Wisconsin},
        postcode={53233},
        country={U.S.A.}}

    \author[inst2]{Anindya Ganguly}
    \author[inst2]{Eirini Goudeli}

    \affiliation[inst2]{organization={Department of Chemical Engineering},
        addressline={University of Melbourne},
        state={Victoria},
        country={Australia}}

    \author[inst3,inst4]{Georgios A. Kelesidis}

	\affiliation[inst3]{organization={Nanoscience and Advanced Materials Center (NAMC), Environmental and Occupational Health Science Institute, School of Public Health, Rutgers, The State University of New Jersey},
		addressline={170 Frelinghuysen Road},
		city={Piscataway},
		state={New Jersey, 08854},
		country={U.S.A.}}

    \affiliation[inst4]{organization={Particle Technology Laboratory, Institute of Process Engineering, Department of Mechanical and Process Engineering},
        addressline={ETH Z$\ddot{u}$rich},
        city={Sonneggstrasse 3},
        state={CH-8092 Z$\ddot{u}$rich},
        country={Switzerland}}

	\author[inst1]{Somesh P. Roy\corref{cor1}}
    \cortext[cor1]{Corresponding author: somesh.roy@marquette.edu}

    \begin{abstract}
        Incipient soot particles obtained from a series of reactive molecular dynamics simulations were studied to understand the evolution of physical, chemical, and morphological properties of incipient soot. Reactive molecular dynamics simulations of acetylene pyrolysis were performed using ReaxFF potential at 1350, 1500, 1650, and 1800~K. A total of 3324 incipient soot particles were extracted from the simulations at various stages of development. Features such as the number of carbon and hydrogen atoms, number of ring structures, mass, C/H ratio, radius of gyration, surface area, volume, atomic fractal dimension, and density were calculated for each particle. The calculated values of density and C/H ratio matched well with experimental values reported in the literature.
        Based on the calculated features, the particles were classified in two types: type 1 and type 2 particles. 
        It was found that type 1 particles show significant morphological evolution while type 2 particles undergo chemical restructuring without any significant morphological change. The particle volume was found to be well-correlated with \editsf{the} number of carbon atoms in both type 1 and type 2 particle, whereas surface area was found to be correlated with the number of carbon atoms only for type 1 particles. 
        A correlation matrix comparing the level of correlation between any two features for both type 1 and type 2 particle was created.
        Finally, based on the calculated statistics, a set of correlations among various physical and morphological parameters of incipient soot was proposed.
    \end{abstract}


\end{frontmatter}


\section{Introduction} \label{s:intro}

Soot, also known as black carbon, is a particulate matter that is an unwanted byproduct of incomplete combustion of hydrocarbon fuels \cite{Michelsen2020Oct}. It greatly influences the radiative energy balance of the atmosphere and is a major forcing factor behind climate change~\cite{Hansen2004Jan,Bond2013Jun}. It also impacts public health and welfare and is one of the leading causes of mortality worldwide~\cite{Lim2012}.
Exposure to soot or black carbon can lead to serious health issues such as cancer \cite{Jacobsen2008Jul} and cardiovascular diseases \cite{Jerrett2013Aug}.  Due to the ubiquitous presence of combustion events -- both natural and anthropogenic -- it is essential to understand the formation and evolution of soot particles in order to regulate and control their detrimental effects~\cite{soot_importance}.

The precise mechanism underlying the formation of soot particulates from gaseous species remains uncertain, primarily due to the intricate chemical composition of the hydrocarbon system, the multiphysics interplay, and the multiscale nature of the soot formation process. The soot formation and growth involve a series of complex physical and chemical processes~\cite{Irimiea2019Apr,Appel2000Apr}. These include the generation of gaseous precursor molecules such as polycyclic aromatic hydrocarbons (PAHs)\cite{Dobbins1994, Balthasar2003May, Wang2016Feb}, the inception of the incipient soot particles through the physical and chemical interactions among these precursor molecules~\cite{Michelsen2020Oct, Wang2011Jan}, the \editsf{aging} of these particles via surface growth due to dehydrogenation and carbon-addition from the gas phase and due to aggregation via coagulation and coalescence~\cite{Michelsen2020Oct,Johansson2017Dec,Rigopoulos2019}, and the fragmentation and oxidation of soot particles~\cite{Michelsen2020Oct,Wang2011Jan,Rigopoulos2019,Russo2015Jan}.

Due to the complex physicochemical interactions, soot undergoes significant alterations in its internal structure and physical properties during its evolution. These modifications occur as the particles progress from the initial stage of incipient soot to the intermediate stage of young soot, ultimately culminating in the final stage of mature soot~\cite{Michelsen2020Oct}. 
Initially, clusters of gaseous PAHs come together to form the first incipient soot particles. During the later stages of formation, the incipient soot undergoes surface chemical reactions and surface condensation of PAHs, resulting in the transformation of incipient soot into young soot. The properties of young soot can be described as having a liquid-like structure and a relatively low level of graphitization \cite{Michelsen2020Oct,Baldelli2020Aug,Zhou2023Jan}. Mature soot exhibits a highly organized graphitic structure that possesses a minimal level of curvature \cite{Baldelli2020Aug,Zhou2023Jan}.

\editsf{The C/H ratio of soot particles increases as the particles get matured. For example,  the C/H ratio of freshly formed soot particles is less than 2.0 \cite{Schulz2019Jan} while the young soot particles can attain a C/H ratio ranging from 2.0 to 4.5 \cite{Minutolo1996Jan}. Mature soot particles usually have a C/H ratio of 4.5 or higher\cite{Minutolo1996Jan}.The average C/H ratio of soot particles from a premixed ethylene flame was also reported by Schulz et al. \cite{Schulz2019Jan} to be around 2.33.
The density of the soot particles also changes as the particles go through maturation.  The \editsf{empirical} density of young and mature soot particle is measured to be 1.5 g/cm$^3$ \cite{Camacho2015Oct} and 1.7-1.9 g/cm$^3$ \cite{Park2004Sep}, respectively. The values reported or assumed in multiple other studies~\cite{Zhao2007Jan, Veshkini2015Nov, desgroux:hal-03326326,Johansson2017Dec} are comparable to this.}

The internal structure of soot consists of both aromatic and aliphatic carbon atoms. The presence of 5- and 6-member ring structures have been identified as an important feature of the soot core~\cite{Johansson2017Dec, Semenikhin2021, Frenklach2020, Frenklach2021}, which influences mechanical properties of soot~\cite{Pascazio2020Sep}.
The physicochemical properties of mature soot particle that come out of a combustion system is strongly dependent on the internal structure of particles, which in turn depend on the physicochemical evolution process that the particle has gone through since its formation.~\cite{Michelsen2020Oct}. 

In recent years, there have been remarkable experimental advances into understanding the internal structure of soot.  Almost all of these recent experimental findings show the importance of ring structures in various stages of soot. For example, Schulz et al.~\cite{Schulz2019Jan} observed multiple aromatic compounds with aliphatic side chains; Gleason et al.~\cite{Gleason2021Jan} showed that soot nuclei can form from aromatic compounds with only one or two rings; Cheng et al.~\cite{Chang2020Jun} observed structural changes such as onset of micropores and graphitic microcrystals of carbon black and soot particles during their evolution; Carbone et al.~\cite{Carbone2017Jul} found that the optical \editsf{properties} of soot \editsf{change} as soot graphitizes during its evolution; Jacobson et al.~\cite{Jacobson2020Mar} explored the molecular structure of soot and concentration of PAH in soot; Commodo et al. provided insights of the importance of $sp^3$ carbon and advanced graphitic structure \cite{Commodo2017Jul} and aliphatic pentagonal rings~\cite{Commodo2019Jul} in the early stages of soot formation. \editsf{When analyzing the radiation scattering by soot particles using refractive index, discrete element modeling with discrete dipole approximation simulations showed that the mass absorption coefficient of soot increases by up to 75\% with increasing residence duration in premixed fires of low equivalency ratio \cite{Kelesidis2019Jan}. Increased soot maturity is the cause of this, which is in great agreement with evidence obtained from laser induced incandescence in ethylene flames \cite{Olofsson2015Jun} and premixed methane \cite{Bejaoui2015Mar}.}

However, despite these recent findings, our knowledge about the evolution of internal and physicochemical properties of soot is still incomplete. Since the exact physical and chemical pathways behind the formation and evolution of soot particles from incipient to mature are still not fully known, there is considerable uncertainty in the estimation of the physical, chemical, and morphological properties of soot at various stages. This poses difficulty down the line in engineering-scale modeling, where the goal is often to model soot emissions from combustion systems at a scale relevant to real-world devices. Engineering-scale reactive CFD models are computational models that operate at the continuum scale and are designed to capture the continuum-scale dynamics of reactive flow. These models are capable of replicating the combustion behavior of real-world systems, including laboratory-scale flames, internal combustion engines, and fires, and can provide estimations of both local and global characteristics of the reactive flow field. 
Because of the inherent complexity involved in combustion modeling and the vast difference in scales, it is impractical for engineering-scale models to keep track of the evolution of the internal atomic structures of the particle.
Therefore, engineering-scale models of soot formation often depend on approximations. Notable numerical soot models such as the semi-empirical two-equation model \cite{Leung1991Dec}, method of moments \cite{Frenklach2002May,Roy2014Apr,Balthasar2005Jan,mbp_hmom_09,mf_05}, stochastic soot model \cite{ Mosbach2009Jun}, discrete sectional model~\cite{Kalbhor2023Sep,Huo2021Aug,gts_80, wf_88, RoyCST2016}, etc. all contain several approximations and/or hypotheses to simplify the inception and growth of soot. These approximations, which include, but are not limited to, inception due to  dimerization of soot precursor \cite{Frenklach2002May,Roy2014Apr}, the spherical shape of  particles~\cite{Frenklach2002May,Leung1991Dec}, constant density \cite{Mosbach2009Jun}, etc. exist partly due to our incomplete knowledge of underlying soot-related processes and partly due to the complexity of combustion modeling.

In this work, we attempt to mitigate some of these shortcomings of engineering-scale soot models by providing an improved estimation -- by way of reactive molecular dynamics simulation -- of various soot properties as soot evolves in combustion systems. We extend our previous work on acetylene pyrolysis using reactive molecular dynamics (RMD) at four different process temperatures~\cite{Mukut_paper1} to investigate how physical, chemical, and morphological properties of incipient soot evolve and how are they correlated with one another.

Reactive molecular dynamics allows \editsf{for modeling} the chemical evolution in a reacting system by tracking bond breakage and bond formation among atoms. This provides an unprecedented view of physicochemical transformations that take place during soot inception~\cite{Schuetz2002Jan,Mao2017Sep,Han2017Aug,Chen2020Sep}. One of the most common tools for modeling reactive hydrocarbon systems in RMD is the reactive force field (ReaxFF)~\cite{vanDuin2001Oct} for carbon, hydrogen, and oxygen chemistry~\cite{Ashraf2017Feb,Chenoweth2008Feb}.
RMD has been used to explore pyrene dimerization~\cite{Schuetz2002Jan}, which is often treated as a key step in soot nucleation. RMD has also been utilized to look at other mechanism of soot nanoparticle formation~\cite{Han2017Aug}, the inception of soot from different PAHs such as naphthalene, pyrene, coronene, ovalene and circumcoronene~\cite{Mao2017Sep}. Recently, the effect of oxygenated additives on diesel soot was also explored via RMD~\cite{Chen2020Sep}.
RMD also allows for a detailed atomic exploration of the internal structure of soot~\cite{Pascazio2020Sep, Mukut_paper1}. For example, the amount of cross-linking in the core and shell structure of developing and mature soot particles were explored in~\cite{Pascazio2020Sep}. In our previous study, we also identified the presence of a denser core with \editsf{the existence of ring structures} at the center of larger incipient soot particles~\cite{Mukut_paper1}.

This study presents findings from a set of isothermal RMD simulations of acetylene pyrolysis at 1350, 1500, 1650, and 1800 K using the ReaxFF potential. We study four different process temperature because some researchers have indicated the influence of temperature on soot microstructure, surface reactivity, and degree of graphitization\cite{Pathak2022Apr, Liu2017Mar}.
The RMD simulations at the four temperatures yield numerous incipient soot particles at different stages of growth. The physicochemical characteristics of the soot particles  are subsequently examined to explore the statistical measures that can be used to describe the development of the initial soot particle. The main objective of this article is to provide insights into the physicochemical and morphological characteristics of early-stage soot particles acquired through RMD simulations and to explore possible correlations among various properties which can potentially be used to improve engineering-scale soot models.

\section{Numerical Methodology} \label{s:numerics}
\subsection{Simulation configurations}
The RMD approach taken in this study has been reported in~\cite{Sharma2021Aug}, and the exact simulation configurations for this study have been previously described in~\cite{Mukut_paper1}. Therefore, we only provide a brief summary here.
1000 acetylene molecules are randomly distributed within a $75 \text{\AA} \times 75\text{\AA} \times 75 \text{\AA}$ cubic domain at four different temperatures: 1350, 1500, 1650, and 1800~K. \editsf{This temperature range is consistent with those used in laminar flames \cite{Zhao2005Jan} or flow reactors \cite{Mei2021Apr}.} Each configuration \editsf{was} simulated multiple times with different randomization. The simulations were done in Large-scale Atomic/Molecular Massively Parallel Simulator (LAMMPS) software \cite{Thompson2022Feb} using the ReaxFF potential~\cite{vanDuin2001Oct, Castro-Marcano2012Mar} with  a timestep of 0.25~fs following Mao et al.~\cite{Mao2017Sep}.
The simulations are conducted using the NVT (constant number, volume, and temperature) ensemble using the velocity-Verlet algorithm~\cite{Swope1982Jan} along with the Nose-Hoover thermostat~\cite{Evans1985Oct}.  The simulation results are examined at intervals of \editsf{0.05~ns}, and large hydrocarbon clusters are identified, organized, and analyzed using a variety of tools including MSMS~\cite{Sanner1996Mar}, MAFIA-MD~\cite{Mukut2022Jul}, and OVITO \cite{Stukowski2009Dec}.

\begin{figure} [!htbp]
    \centering
    \includegraphics[width=\linewidth]{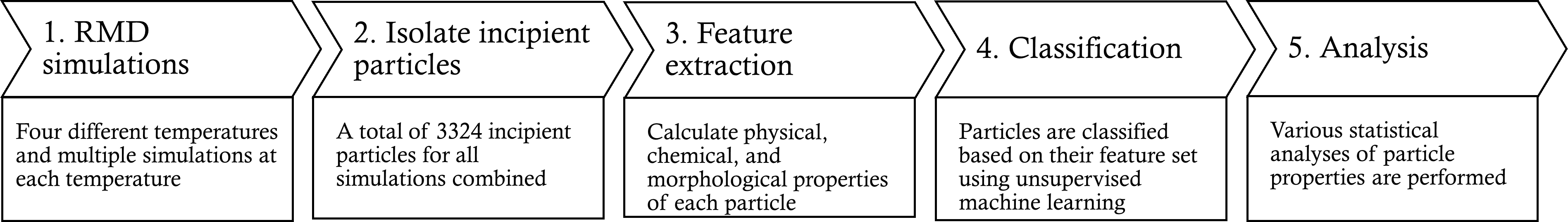}
    \caption{Overview of the workflow employed in this study. The blocks numbered (1) to (4) have been discussed in more detail in our previous work~\cite{Mukut_paper1}. This work focuses on the block (5).}
    \label{f:workflow}
\end{figure}

The overall workflow for this work is depicted in Fig.~\ref{f:workflow}. A total of 24 RMD simulations are performed at four different temperatures. From each simulation, the large molecular clusters are identified as potential incipient soot at every \editsf{0.05~ns}. These clusters are molecules having more than 20 carbon atoms and at least one 5-, 6-, or 7-member ring structure (indicating the possible presence of an aromatic component). These criteria are based on previous findings~\cite{Mukut2021May}. Incidentally, in our simulations, the smallest molecular cluster matching these criteria was found to have 65 carbon atoms. A total of 3324 soot particles are identified at various stages of their development from the 24 simulations. Then a variety of physical (e.g., density and mass), chemical (e.g., number of rings), and morphological (e.g., volume and surface area) features of each particle are quantified. These features are described in Sec.~\ref{s:physicochemical}. Following this feature extraction, the particles are classified using two unsupervised machine-learning techniques: k-means clustering \cite{Lloyd1982Mar} and t-distributed stochastic neighbor embedding (t-SNE) \cite{vanderMaaten2008}. As discussed in our previous work~\cite{Mukut_paper1}, this classification results in two distinct classes of incipient particles, referred to as Type 1 and Type 2 incipient particles.
Finally, various statistics including the correlation among different properties are evaluated for the entire set of particles. The correlation between the physicochemical
properties are estimated using Kendall's Tau correlation \cite{Kendall1938Jun}. Kendall's Tau is a non-parametric measure of the relationship between columns of ranked data. It estimates a correlation coefficient ($\tau$) with a value between $-1$ and $+1$, where $+1$ means perfect positive correlation, $-1$ means perfect negative correlation, and 0 means no correlation between variables. It also provides a p-value that indicates the statistical significance of the correlation.

\subsection{Extraction of physicochemical properties} \label{s:physicochemical}
Once the soot particles are identified and isolated using the cluster analysis tool from the OVITO python module \cite{Stukowski2009Dec}, the physicochemical properties of each identified soot particle are evaluated using in-house Python script and other tools such as MSMS~\cite{Sanner1996Mar} and MAFIA-MD~\cite{Mukut2022Jul}. 
Key features such as number of atoms ($N$), carbon to hydrogen ratio ($\chr$), and particle mass ($M_p$) can be obtained directly from the output log of RMD simulation (a trajectory file). The solvent-excluded volume ($V$) and solvent-excluded surface area ($A$) are calculated using MSMS~\cite{Sanner1996Mar} using a probe radius of 1.5\AA. Some other features are calculated by simple algebraic, or geometric analysis or by using correlations proposed in the literature as discussed below.

The radius of gyration ($R_g$) is calculated geometrically using the atomic coordinates extracted from the trajectory files using Eqn.~\ref{e:rg}:
\begin{equation}
    \label{e:rg}
    R_g = \sqrt{\frac{\sum_{i=1}^N m_{p,i} r_i^2}{\sum_{i=1}^N m_{p,i}}},
\end{equation}
where $r_i$ is the distance of the $i\textsuperscript{th}$ atom from the center of mass of the molecular cluster, $m_{p,i}$ is the mass of $i\textsuperscript{th}$ atom, and $N$ is the total number of atoms in the cluster.
In engineering-scale soot models, it is often difficult to capture the radius of gyration of incipient particles due to the lack of information on particle morphology and size distribution. However, the mass and volume of the particles are comparatively easier to obtain. Using these, the volume-equivalent radius ($R_{eq}$) can be calculated directly from particle volume from Eqn.~\ref{eq:R_eq}:
\begin{equation} \label{eq:R_eq}
    R_{eq} = \left(\frac{3V}{4\pi}\right)^{\nicefrac{1}{3}}.
\end{equation}

The atomic fractal dimension ($D_f$) is calculated using the sandbox method \cite{Theiler1990Jun,Forrest1979May} using Eqn.~\ref{e:df}.
\begin{equation}
    \label{e:df}
    D_f = \frac{\log M_p(r)}{\log r},
\end{equation}
where $M_p(r)$ is the mass of atoms in the cluster as a function of radial distance from center of mass ($r$).
Atomic fractal dimension ($D_f$) is, therefore, the slope of the log-log plot
of $M_p(r)$ vs. $r$. As discussed in~\cite{Sharma2021Aug}, here the fractal dimension is calculated using the existing atoms in an incipient particle. Hence, the term ``atomic'' fractal dimension is used to remove any confusion with the fractal dimension of aggregates which is often used to calculate the level of geometric self-similarity in soot
aggregates \cite{Suarez1264229,Wang2022Jul, Wang2017Nov}. An atomic fractal dimension of 1 represents a linear structure while a value of 3 indicates a perfectly spherical shape \cite{Wang2022Jul}.

The \editsf{simulated primary particle density (shortened as simulated density, or $\rho_{s}$)} of the incipient soot particles is obtained from the mass and volume of the incipient particle (Eqn.~\ref{e:actual-denstiy}): 
\begin{equation}
    \rho_{s} = \frac{M_p}{V}\label{e:actual-denstiy}
\end{equation}
We also calculate a widely used metric -- empirical density ($\rho_{e}$) \editsf{also referred to as bulk density in the contemporary literatures},  which can convey additional information about the maturity of the incipient particles. Empirical density of an incipient particle changes with the level of maturity of the particle and is calculated using Eqn.~\ref{e:bulk-density}~\cite{Johansson2017Dec, DeCarlo2004Jan}:
\begin{eqnarray}
    \rho_{e} &= (0.260884 a^2c)^{-1}\left(\frac{w_C \chr +w_H}{\chr +1}\right),    \label{e:bulk-density}
\end{eqnarray}
where $w_C$ is the weight of carbon atoms (12.011 g/mol), $w_H$ is the weight of
hydrogen atoms (1.008 g/mol), $a$ is the length of the graphite unit cell in
the basal plane (2.46 \AA), $c$ is the interlayer spacing in Angstroms (3.50~\AA~for soot), and $\chr$ represents the carbon to hydrogen ratio of the cluster.

Finally, \texttt{MAFIA-MD}~\cite{Mukut2022Jul} code is used to identify 5-~/6-~/7-membered ring structures in a molecular cluster. As discussed in detail in~\cite{Mukut2022Jul}, \texttt{MAFIA-MD} does not strictly check for Huckel's rules of aromaticity, and hence, to remove any confusion, we will use the
terms ``ring'' or ``cyclic'' in this manuscript instead of aromatic when discussing these internal structures in the soot clusters. The number of 5-~/6-~/7-membered rings are denoted as $\Nfive$, $\Nsix$, $\Nseven$, respectively, and the total number of rings is denoted as $\Nring$. Similarly, number of carbons in rings are denoted as $\Ncring$ and the number of non-cyclic carbons in a particle is denoted as $\Ncnonring$.

\section{Results and Discussion} \label{s:resultDiscussion}

\subsection{Inception of soot and classification of incipient soot particles}
\label{ss:incipientSoot}
The general sequence of events as seen from the RMD simulations leading up to the inception events have been discussed in detail in our earlier works~\cite{Mukut2021May, Mukut_paper1}.
Initially, the acetylene molecules combine to form small linear chains. Subsequently, these linear chains undergo cyclization, transforming into cyclic structures. Following the process of cyclization, the small clusters start to grow through surface reactions that involve bond formation, as well as internal reorganization leading to incipient soot particles~\cite{Mukut2021May, Mukut_paper1}. A similar process has been also been reported by Zhang et al.~\cite{Zhang2015Apr} and Sharma et al. \cite{Sharma2021Aug}.

\begin{figure} [!htbp]
    \centering
    \includegraphics[width=0.6\linewidth]{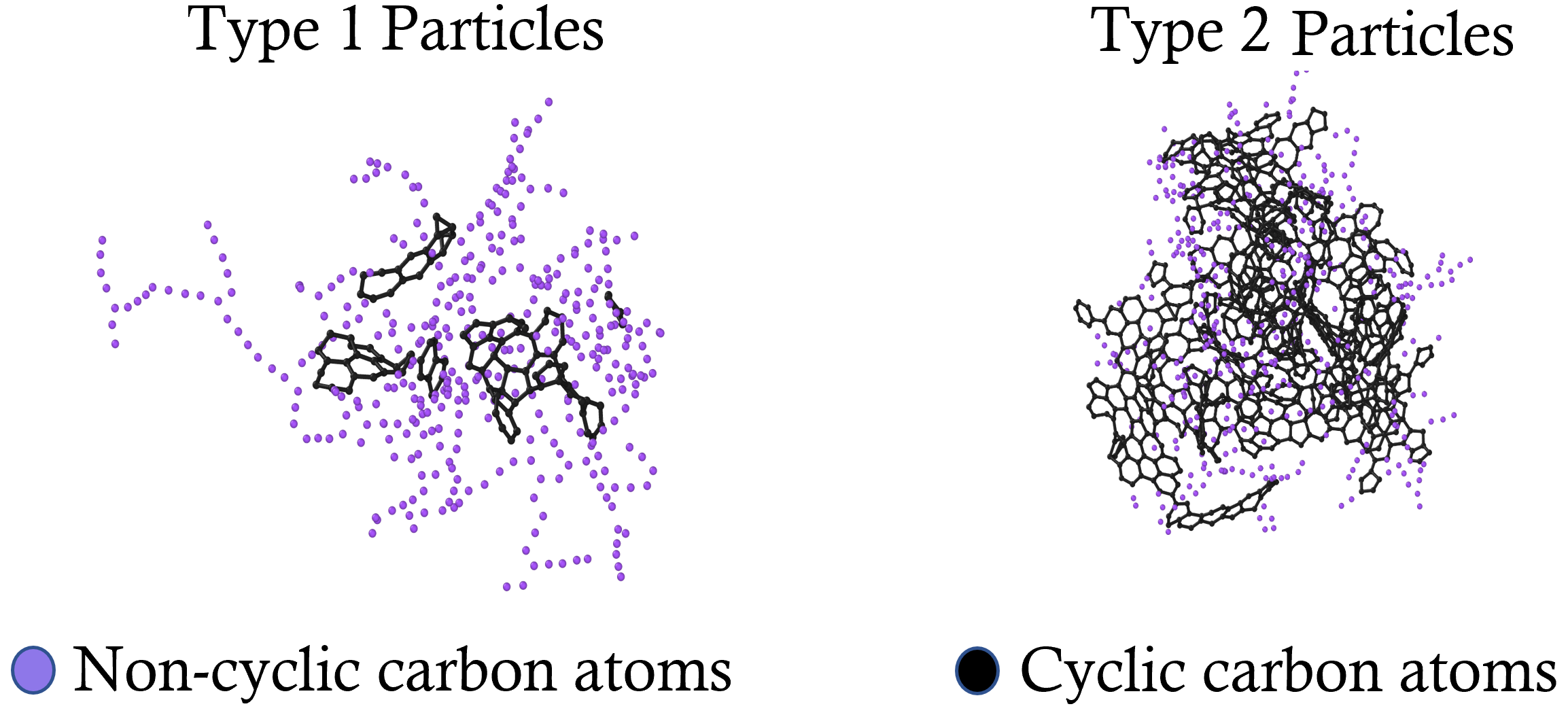}
    \caption{One example particle from each class obtained from RMD simulations.}
    \label{f:sampleParticles}
\end{figure}

\editsf{In the present study, being consistent with the current understanding of soot particles,} an incipient soot particle is identified as a molecular cluster with more than 20 carbon atoms and with at least one ring structure (5-, 6-, or 7-member rings) \editsf{\cite{Mukut2021May}}. The first such particle in our simulation was found to have 65 carbon atoms. A total of 3324 incipient soot particles, with carbon atoms ranging from 65 to 1503, were extracted at various stages of evolution from all the simulations at four temperatures. These particles showed significant variation in their physicochemical features and were therefore classified into two classes based on all the extracted features using unsupervised machine learning techniques (k-means clustering and t-SNE diagram) as discussed in~\cite{Mukut_paper1}.
For easier identification, the two classes of particles are denoted as ``Type 1'' and ``Type 2'' particles \editsf{(see also Fig. \ref{f:sampleParticles})}. Type 1 particles exhibit a lower number of carbon atoms ($65-818$), whereas, in Type 2 particles, the number of carbon atoms is higher ($759-1503$). This observation highlights the fact that the physicochemical properties of the incipient particles undergo a transition once a certain level of growth is reached~\cite{Mukut_paper1}. It is important to mention that the classification is based on \textit{all physicochemical features} and not based only on the number of carbon atoms. The number of carbon atoms, conveniently, serves as a good approximate indicator for the boundary between type 1 and type 2 particles. Some small type 2 particles may have a lower number of carbon atoms than large type 1 particles (\editsf{there is a} slight overlap in the range of the number of atoms between the two types). In our simulation, we identified a total of 670 type 1 and 2654 type 2 incipient particles. Figure~\ref{f:sampleParticles} shows one sample Type 1 and one sample Type 2 particle. In this representation, the non-cyclic carbon atom structures are denoted by blue dots, while the cyclic structures are denoted by black dots.

It can be observed in Fig. \ref{f:sampleParticles} that there are discernible differences in the internal structure between type 1 and type 2 particles. Type 1 particles possess numerous small aromatic islands embedded within a plentiful network of aliphatic compounds. As the particles transitions from type 1 to type 2, the smaller islands amalgamate into larger aromatic islands situated in the central region of the particles (growth centers \cite{Michelsen2020Oct}), resulting in a decrease in the proportion of aliphatics. In this manner, the liquid-like or young soot particles that are rich in aliphatic compounds (i.e., type 1 particles) transform and develop into partially graphitized soot particles (i.e., type 2 particles) with higher fraction of aromatic components. During this transformation, the soot particles acquire well-defined growth centers. Therefore, it is expected that the intercorrelations among various physicochemical properties will change as the particle transitions from type 1 to type 2.

\subsection{Physicochemical features of incipient particles} \label{ss:physicochemical-results}
The physicochemical features of
soot particles investigated include mass, number of atoms, atomic fractal
dimension, volume, surface area, density, particle radius, and statistics of
cyclic structures. The physical properties (for example, mass, volume, density,
etc.) are often the features that are used in engineering-scale soot models for
reactive CFD \editsf{as well as in discrete element modeling \cite{Kelesidis2017Jan} and dipole approximation \cite{Kelesidis2019Nov, Kelesidis2019Jan}}. A better understanding of these features and their inter-correlation will help improve these engineering-scale soot models. The
chemical properties such as the number of atoms, C/H ratio and statistics of
cyclic/non-cyclic structures are important metrics by which the evolution and
maturity of soot particles can be tracked. \editsf{The evolution of the soot C/H ratio is essential to estimate the evolution of its optical properties \cite{Kelesidis2021Feb}  and assist its detection by laser diagnostics \cite{Kelesidis2021Feb,Kelesidis2022Jun}.}

In this work, we focused on the analysis for a set of select physical (mass,
radius, surface area, and volume) and chemical (statistics of ring structures)
features as the particle grows. As mentioned in Sec.~\ref{ss:incipientSoot}, as the
particle grows, it transitions from type 1 to type 2. Since the demarcation of
these types can be very closely tracked by the number of carbon atoms in the particle,
which correlates very well with the mass and size of the particle, we track the
growth of a particle via either the total number of carbon atoms or the particle mass or the particle radius. The Kendall's Tau~\cite{Kendall1938Jun} statistical test is performed to determine the level of correlation between the different properties. The
correlation coefficients ($\tau$) of the Kendall's Tau statistical test are
reported in each figure. It was found that whenever a good correlation (i.e.,$|\tau|\gg0$) was found between two variables, the test yielded a very
small p-value (almost zero), indicating a high statistical significance.
Therefore, the p-value is not reported in the figures for brevity. We
visualize the data using a joint plot of scatter plot and
histogram/distribution using the python package \texttt{seaborn}
\cite{Waskom2017Jul}.

\begin{figure}[!htbp]
    \centering
     \includegraphics[width=\linewidth]{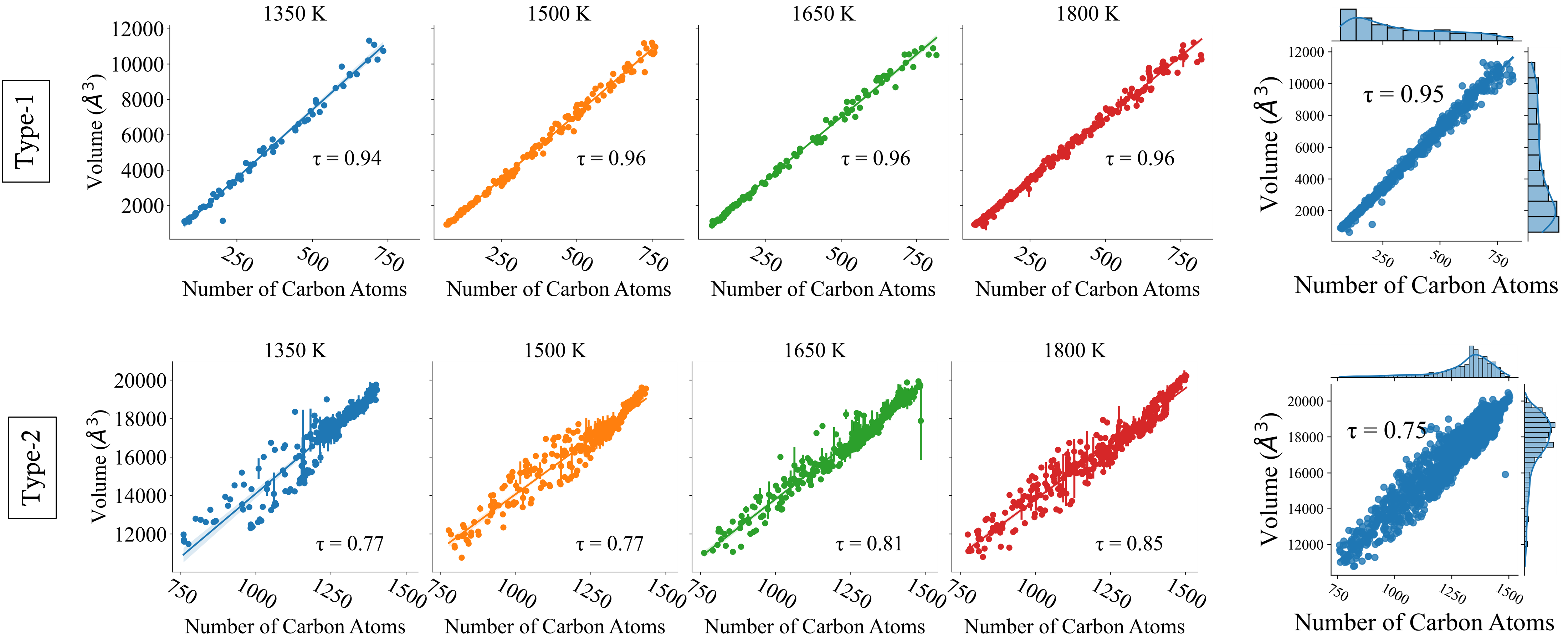}
    \caption{Evolution of particle volume with the number of carbon atoms at different temperatures. Top row: Type~1 particles, bottom row: Type~2 particles.} \label{f:T_Vol} 
\end{figure}

\begin{figure}[!htbp]
    \centering
     \includegraphics[width=\linewidth]{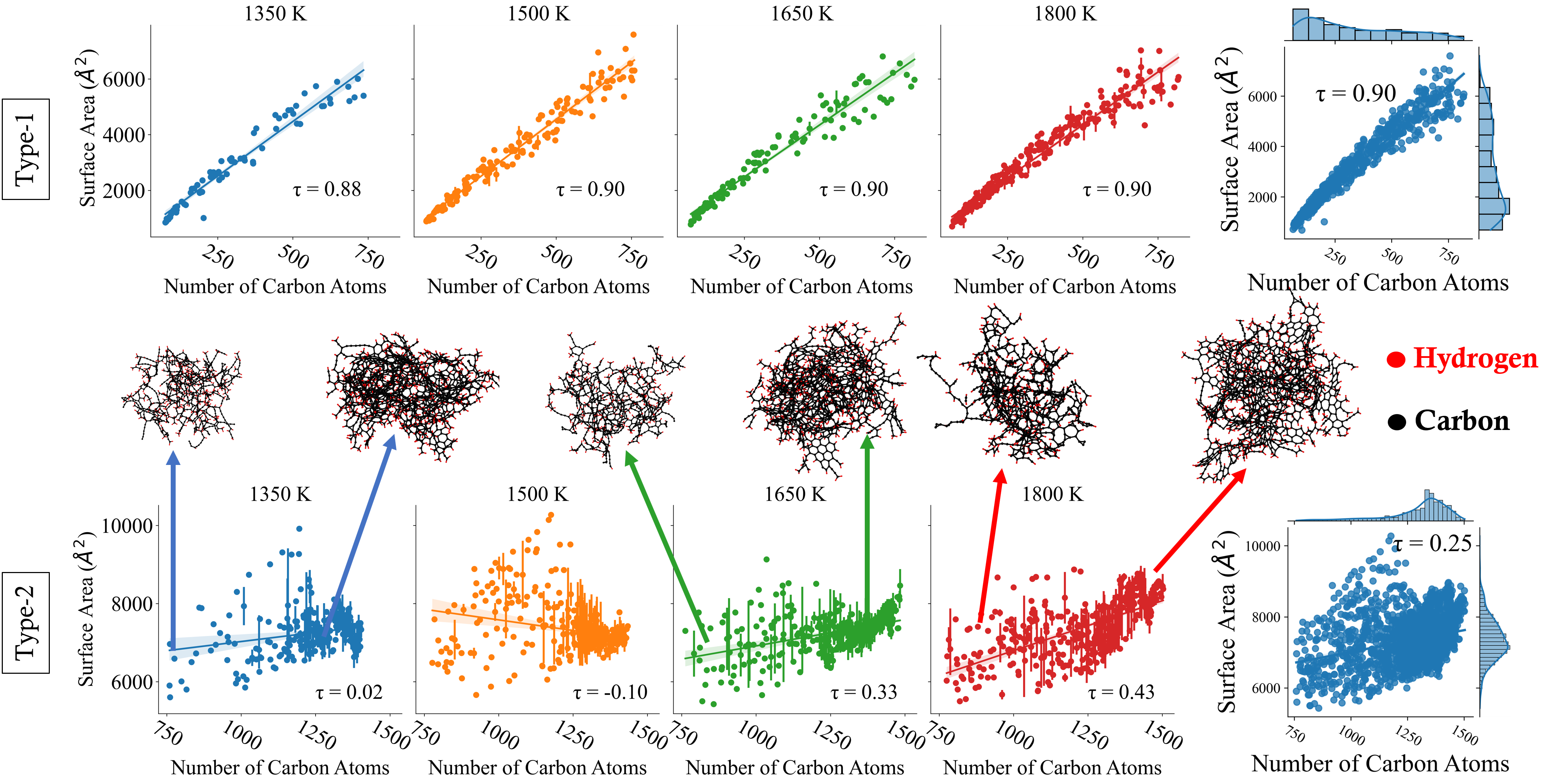}
    \caption{Evolution of particle surface area with the number of carbon atoms at different temperatures. Top row: Type~1 particles, bottom row: Type~2 particles.} \label{f:T_Area}
\end{figure}

The evolution of particle volume and surface area is presented in Figs.~\ref{f:T_Vol} and~\ref{f:T_Area} for both type 1
(top row) and type 2 particles (bottom row) as a function of the number of
carbon atoms in the particles. The first four columns of plots correspond to simulation
temperatures of 1350, 1500, 1650, and 1800 K, whereas the rightmost  column combines all
the temperatures together. \editsf{A trendline is also shown on the scatter plots for each temperature.} From Fig.~\ref{f:T_Vol}, we see that both type 1
and type 2 particles have an excellent correlation between the particle volume and
the number of carbon atoms in the particles across all temperatures. The correlation
coefficient ($\tau$) is very close to 1 for type 1 particles, representing
an almost perfect correlation between the variables. For type 2 particles, the
correlation coefficient ($\tau$) is lower, but still close to~1. This
indicates that the particle volume is well correlated with the number of
carbon atoms in both type 1 and type 2 particles. Another important observation
is that the particle volume and number of carbon atoms seem to be correlated in the
same way for all simulation temperatures. 
The slopes of the lines that provide linear fit between the volume and the number of carbon atoms do not vary much with temperature (discussed further later in Table~\ref{tab:surface_volume_relation_temp}). This shows that the relationship between particle volume and number of carbon atoms is not affected by the temperature.

\begin{figure} [!htb]
    \centering
    \includegraphics[width=0.6\linewidth]{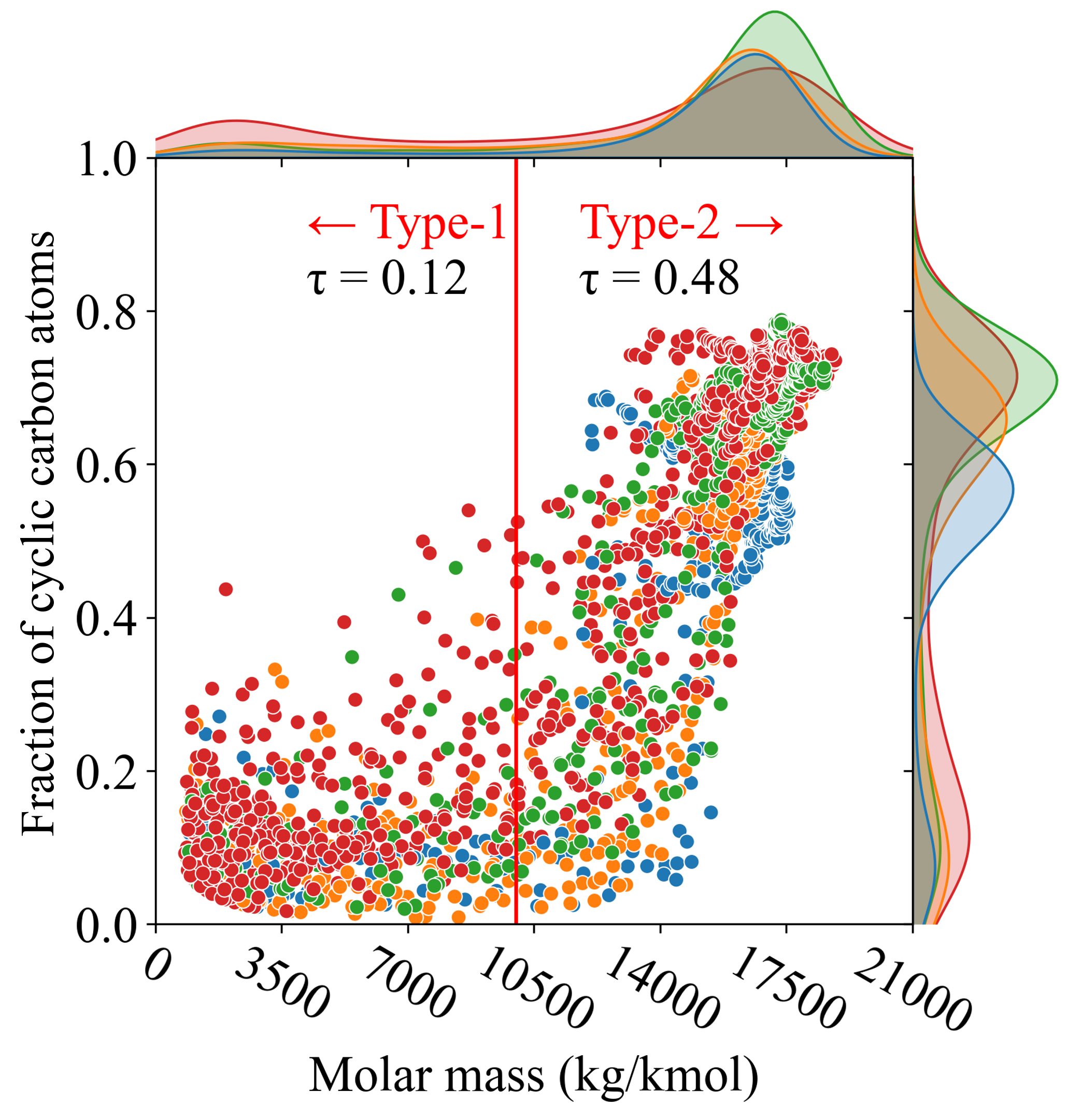}
    \caption{Evolution of fraction of carbon atoms in ring structures ($\Ncring/\Nc$) with molar mass.} \label{f:M_cyclic}
\end{figure}

For type 1 particles, a similar conclusion can also be drawn from Fig. \ref{f:T_Area} for the surface area: a very good correlation is observed with the number of carbon atoms. But for the type 2 particles, the surface area does not seem to be correlated with the number of carbon atoms. As the particle transitions from type 1 to type 2, a change from the initial linear growth to the extensive reorganization of the surface is observed. This is shown using some actual particles at the two extremes of the number of carbon atoms for 1350, 1650 and 1800 K in the insets of Fig.~\ref{f:T_Area}. As can be seen, while the topology of the surface shows significant difference between a small type 2 particle ($\Nc\sim800$) and a large type 2 particle ($\Nc\sim1400$), the actual difference in surface area is small.
It shows that the soot surface goes through extensive rearrangement in type 2 particles, which makes the surface area hard to correlate with the number of carbon atoms. The rearrangement of the surface area occurs mostly because of graphitization of the surface as discussed in the internal structure of type 2 particles in~\cite{Mukut_paper1}. The graphitization of the surface is a very complex process and is not well understood. However, it is known that the graphitization of the surface is a slow process, and it depends on the size  of the primary particle itself~\cite{Apicella2019Jun}. We can also note that there is a lower number of ring structures in the similar-sized small type~2 particles ($\Nc\sim800-1000$) in 1800~K than in 1350~K. Temperature can accelerate surface graphitization \cite{kobayashi2018pahs} and, therefore, while comparing the small type 2 particles ($\Nc\sim800-1000$) we observe the formation of graphitized particles (indicated by a large presence of cyclic structures) earlier in 1800~K simulation than in 1350~K simulation. \editsf{The fraction of cyclic carbon atoms is plotted against molar mass in Fig.~\ref{f:M_cyclic}. For type 1 particles, the fraction of cyclic carbon atoms is mostly insensitive to the particle mass. However, for type 2 particles, the fraction of cyclic carbon atoms increases rapidly with molar mass. This further supports the inference regarding the extensive reorganization of the surface in type 2 particles due to graphitization.
This is consistent with results reported by Liu et al.~\cite{LiuJan2017} where it is shown that smaller soot particles  take longer to graphitize compared to the larger ones.} This lack of graphitization leads to a good correlation between surface area and the number of atoms in type 1 particles, which are primarily seen at
the early stage of soot formation. Furthermore, the process temperature does not seem to
affect the quality of correlation or lack thereof.

From Figs.~\ref{f:T_Vol} and~\ref{f:T_Area}, linear trends with the number of carbon atoms are observed for particle volume (type 1 and type 2) and surface area (type 1). A set of linear equations were fitted between particle volume  and surface area with the number of carbon atoms for different temperatures and presented in Table~\ref{tab:surface_volume_relation_temp}. For particle volume, both type 1 and type 2 particles show a very good linear relationship with the number of carbon atoms ($N_C$) as indicated by high $R^2$ values across different temperatures and classes of particles. The $R^2$, or the coefficient of determination  ($R^2\in[0,1]$) is a statistical measure of how close the data are to the fitted regression line. As discussed previously, for surface area, a good correlation is only observed for type 1 particles and therefore, no correlation for surface area for type 2 particles is presented in Table~\ref{tab:surface_volume_relation_temp}. The slope of these linear correlations are very close to one another supporting the observation of the lack of influence of temperature on the quality of the correlation between these quantities. Therefore, we constructed a set of temperature-independent correlations for volume and surface area by aggregating data from all the temperature in Correlation Set~\ref{tab:surface_volume_relation_all}.

\begin{table}[!htbp]
    \caption{Equation of linear curves fitted to the data for different temperatures shown in Figs.~\ref{f:T_Vol} and~\ref{f:T_Area}. 
    $V$ is the particle volume in \AA$^3$,
    $A$ is the particle surface area in \AA$^2$, and
$N_C$ is the total number of carbon atoms.}
    \label{tab:surface_volume_relation_temp}
    \resizebox{\textwidth}{!}{%
    \begin{tabular}{rlllr}
    \hline
    \multicolumn{1}{|r|}{Temperature}      & \multicolumn{1}{r}{Particle Volume (V)}       & \multicolumn{1}{l|}{}                          & Surface Area (A)                             & \multicolumn{1}{l|}{}            \\ \cline{2-5} 
    \multicolumn{1}{|r|}{}                 & \multicolumn{1}{c|}{Type 1}                   & \multicolumn{1}{c|}{Type 2}                    & \multicolumn{1}{c|}{Type 1}                  & \multicolumn{1}{c|}{Type 2}      \\ \hline
    \multicolumn{1}{|r|}{1350 K}           & \multicolumn{1}{l|}{V=15.405 $N_C$ - 297.596} & \multicolumn{1}{l|}{V=13.172 $N_C$ + 930.663}  & \multicolumn{1}{l|}{A=7.842 $N_C$ + 551.652} & \multicolumn{1}{r|}{}            \\
    \multicolumn{1}{|r|}{}                 & \multicolumn{1}{l|}{$R^2$ = 0.987}            & \multicolumn{1}{l|}{$R^2$ = 0.883}             & \multicolumn{1}{l|}{$R^2$ = 0.944}           & \multicolumn{1}{r|}{}            \\ \cline{1-4}
    \multicolumn{1}{|r|}{1500 K}           & \multicolumn{1}{l|}{V=15.405 $N_C$ - 297.596} & \multicolumn{1}{l|}{V=11.41 $N_C$ + 2663.046}  & \multicolumn{1}{l|}{A=8.171 $N_C$ + 464.281} & \multicolumn{1}{r|}{}            \\
    \multicolumn{1}{|r|}{}                 & \multicolumn{1}{l|}{$R^2$ = 0.987}            & \multicolumn{1}{l|}{$R^2$ = 0.913}             & \multicolumn{1}{l|}{$R^2$ = 0.960}           & \multicolumn{1}{r|}{No}          \\ \cline{1-4}
    \multicolumn{1}{|r|}{1650 K}           & \multicolumn{1}{l|}{V=14.207 $N_C$ - 113.638} & \multicolumn{1}{l|}{V=12.028 $N_C$ + 1784.325} & \multicolumn{1}{l|}{A=7.459 $N_C$ + 614.673} & \multicolumn{1}{r|}{meaningful}  \\
    \multicolumn{1}{|r|}{}                 & \multicolumn{1}{l|}{$R^2$ = 0.994}            & \multicolumn{1}{l|}{$R^2$ = 0.934}             & \multicolumn{1}{l|}{$R^2$ = 0.944}           & \multicolumn{1}{r|}{correlation} \\ \cline{1-4}
    \multicolumn{1}{|r|}{1800 K}           & \multicolumn{1}{l|}{V=14.014 $N_C$ -65.388}   & \multicolumn{1}{l|}{V=11.564 $N_C$+2228.574}   & \multicolumn{1}{l|}{A=7.583 $N_C$ + 560.486} & \multicolumn{1}{r|}{}            \\
    \multicolumn{1}{|r|}{}                 & \multicolumn{1}{l|}{$R^2$ = 0.994}            & \multicolumn{1}{l|}{$R^2$ = 0.926}             & \multicolumn{1}{l|}{$R^2$ = 0.953}           & \multicolumn{1}{r|}{}            \\ 
    \hline
    \end{tabular}%
    }
    \end{table}

\noindent
\begin{minipage}{\linewidth} 
\vspace{0.5\baselineskip}
\captionof{Cor}{Equation of curves fitted to the temperature-aggregated data shown in Figs.~\ref{f:T_Vol} and~\ref{f:T_Area}. 
    $V$ is the particle volume in \AA$^3$,
    $A$ is the particle surface area in \AA$^2$, and
$N_C$ is the total number of carbon atoms.}
\label{tab:surface_volume_relation_all}
\textbf{Type 1 particles:}
\begin{eqnarray}    
    V&=14.353 N_C - 119.071, \qquad R^2 = 0.992\\
    A&=7.733 N_C + 547.914, \qquad R^2 = 0.951
\end{eqnarray}

\textbf{Type 2 particles:}
\begin{eqnarray}    
    V&=11.825 N_C + 2151.314, \qquad R^2 = 0.897
\end{eqnarray}    
\vspace{0.5\baselineskip}
\end{minipage}

While we only show the results for particle volume and surface
area, a similar conclusion about the effect of temperature can be drawn from the
evolution of other morphological features, i.e. radius of gyration, volume equivalent radius, number/fraction of cyclic structures, etc. as well . Discussion on these are omitted in this work for brevity.

\begin{figure}[!htbp]
    \centering
\includegraphics[width=0.75\linewidth]{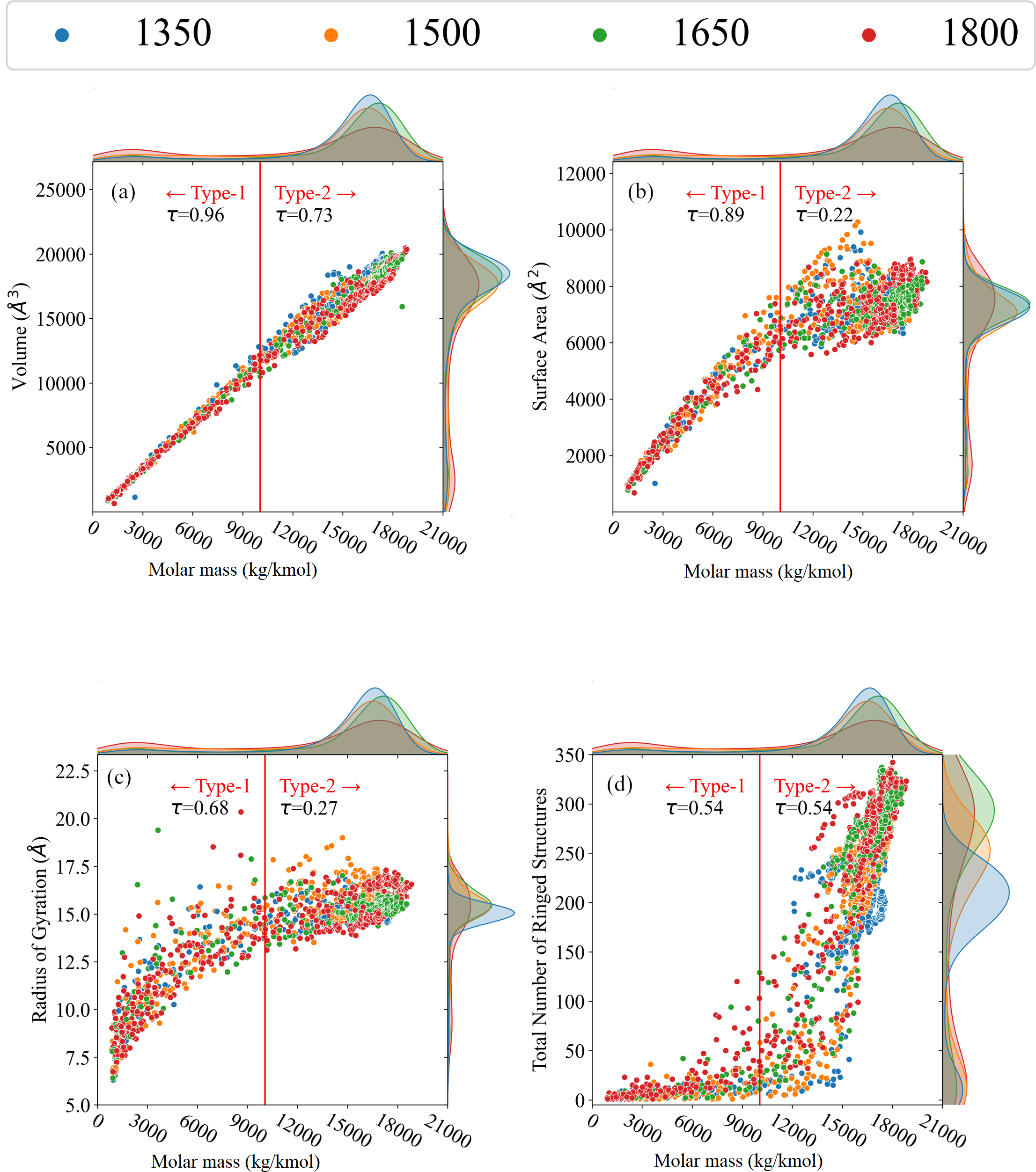}
        \caption{Relationship between molar mass and (a) volume, (b) surface area, (c) radius of gyration, and (d) total number of rings in incipient soot particles}\label{f:Mass_Rel}
\end{figure}

\begin{figure}[!htbp]
    \centering
\includegraphics[width=0.75\linewidth]{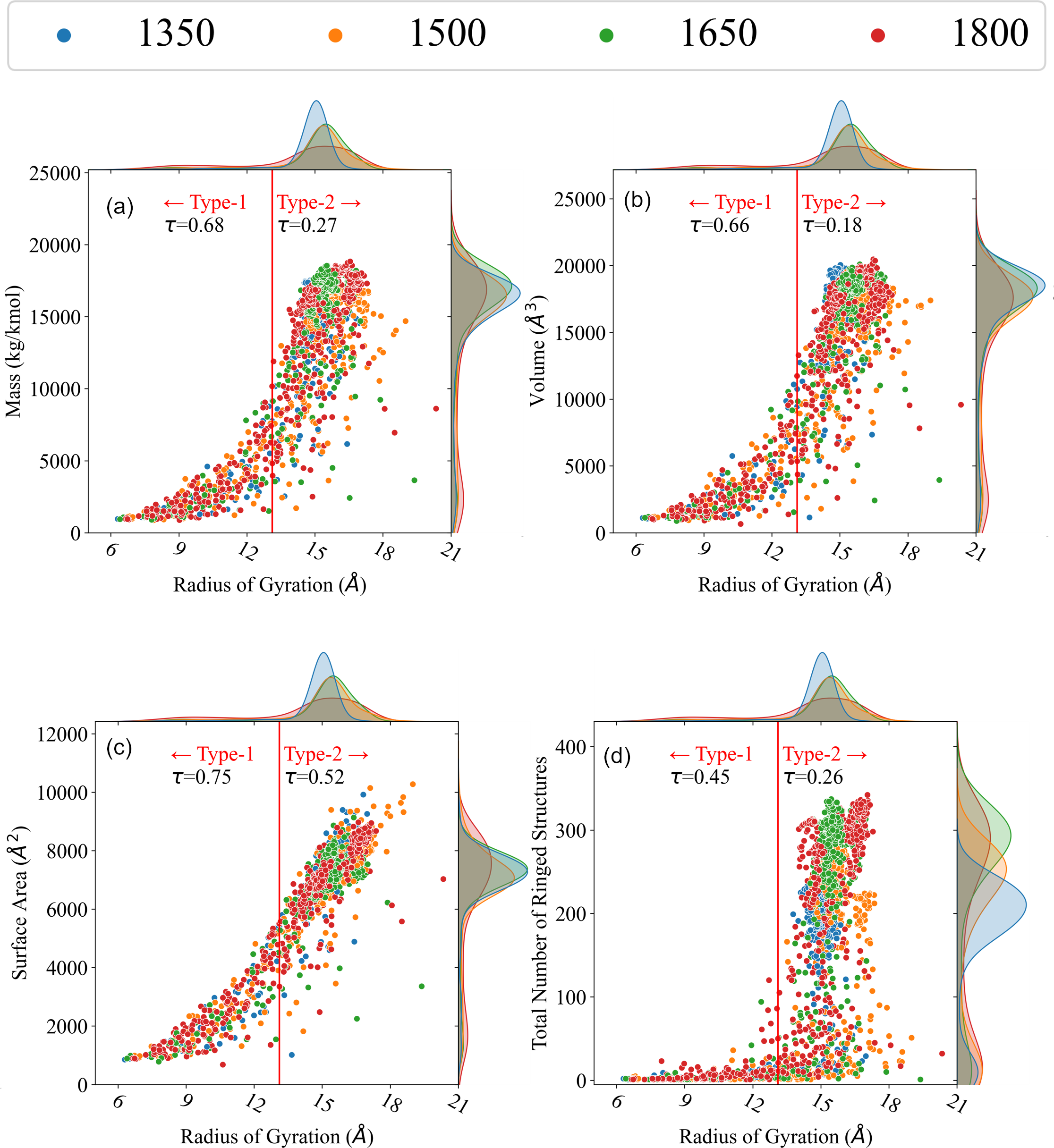}
    \caption{Relationship between radius of gyration and (a) mass, (b) volume, (c) surface area, and (d) total number of rings in incipient soot particles.} \label{f:rg_Rel}
\end{figure}

Most engineering-scale soot models track the evolution of the soot population via the mass or size of soot particles. Therefore, the correlations of particle volume, surface area, and number of rings are shown with molar mass in Fig.~\ref{f:Mass_Rel} and with the radius of gyration in Fig.~\ref{f:rg_Rel}. The results are presented in the form of joint plots which show the distribution of the data points in the form of a scatter plot and the distribution of the individual properties in the form of a density plot along with the temperature markers.
The vertical red line in each plot indicates the boundary between type 1 and type 2 particles and the correlation coefficients ($\tau$) are reported separately for both types of
particles on their respective zones. The transition of a particle from type 1 to type 2 occurs around \editsf{a molar mass of } 10050~kg/kmol or \editsf{a radius of gyration of } 13.8 \AA, and the vertical red lines are drawn at
those locations in these plots. The type 1 particles are on the left of this
line and type 2 is on the right.

The volume, surface area, and radius of gyration of the type 1 particle are well correlated with the molar mass of the particles as seen in Fig.~\ref{f:Mass_Rel}. For type 1 particles, volume and surface area show a linear correlation, while the radius of gyration shows a nonlinear correlation with the mass of the particles.
For the total number of ring structures, however, the correlation is not as good as the other three properties. The nature of the correlations between variables remains unaffected by the temperature. The most important observation from Fig.~\ref{f:Mass_Rel} is the difference in the nature of the correlations between type 1 and type 2. In all cases, the data points become more clustered and
the correlations become weaker except for the total number of ring structures in type~2 particles.
A very sharp increase in the total number of ring structures is observed as the
particles transition to type 2 and grow in mass. The total number of rings is also affected by temperature, even though temperature does not affect the other morphological features. A higher number of cyclic structures is observed with the increase in temperature as we can see from the temperature 
markers and density plots on the right side of the plot.
A set of correlation equations by curve-fitting the data shown in Fig.~\ref{f:Mass_Rel} is presented in Corr.~Set~\ref{tab:correl_w_molarmass}.

Similar conclusions can be drawn using the radius of gyration as the independent variable as shown in Fig.~\ref{f:rg_Rel}.
The volume and surface area, and mass all show a good correlation for type 1 particles. For the
total number of ring structures, however, the correlation is not as good as the
other three properties. The nature of the correlations between features remain
unaffected by the temperature. Similar to Fig.~\ref{f:Mass_Rel}, the type 2
particles show comparatively weaker correlations with the radius of gyration
compared to type 1 particles. 
A set of correlation equations obtained by curve-fitting the data shown in Fig.~\ref{f:rg_Rel} is presented in Corr.~Set~\ref{tab:correl_w_rg}.

The nature of the physical and chemical properties of incipient soot changes as the particle grows from type 1 to type 2. Identification of this transition from type 1 and 2 particles is valuable in the context of engineering-scale soot models as it will allow for updating of property correlation in the soot model without analyzing the internal structure. This may lead to a more comprehensive understanding of soot evolution, including the morphology of primary particles. This is further evident in the discussion of ring structures presented in Sec.~\ref{ss:chem_evolution}.

\vspace{\baselineskip}
\noindent\begin{minipage}{\linewidth}
\captionof{Cor}{Equation of curves fitted to the data shown in Fig.~\ref{f:Mass_Rel}. $M$ is the molar mass of the particle in kg/kmol,
$V$ is the particle volume in \AA$^3$,
$A$ is the particle surface area in \AA$^2$,
$R_g$ is the radius of gyration in \AA, and
$\Nring$ is the total number of ring structures in the particle.}\label{tab:correl_w_molarmass}
\textbf{Type 1 particles:}
\begin{eqnarray}
    V&=&1.160 M - 124.597, \qquad R^2=0.993,\\
    A&=&0.625 M - 545.410, \qquad R^2=0.952,\\
    R_g&=&2.947 \ln(M) - 12.487, \qquad R^2=0.717,\\
    \Nring&=&0.004 M - 5.964, \qquad R^2=0.372.
\end{eqnarray}

\textbf{Type 2 particles:}
\begin{eqnarray}
    V&=&0.935 M - 2375.685, \qquad R^2=0.905,\\
    A&=&0.098 M - 5783.961, \qquad R^2=0.090,\\
    R_g&=&1.933 \ln(M) - 3.293, \qquad R^2=0.114,\\
    \Nring&=&0.036 M - 342.906, \qquad R^2=0.704.
\end{eqnarray}
\vspace{0.5\baselineskip}
\end{minipage}

\vspace{\baselineskip}
\noindent\begin{minipage}{\linewidth}
\captionof{Cor}{Equation of curves fitted to the data shown in Fig.~\ref{f:Mass_Rel}. $V$ is the particle volume in \AA$^3$,
$A$ is the particle surface area in \AA$^2$,
$R_g$ is the radius of gyration in \AA, and
$\Nring$ is the total number of ring structures in the particle.}\label{tab:correl_w_rg}
\textbf{Type 1 particles:}
\begin{eqnarray}
    V&=&995.677 R_g - 6699.483, \qquad R^2=0.647,\\
    A&=&596.486 R_g - 3678.870, \qquad R^2=0.767,\\
    \Nring&=&2.945 R_g - 22.272, \qquad R^2=0.162.
\end{eqnarray}

\textbf{Type 2 particles:}
\begin{eqnarray}
    V&=&807.096 R_g + 4944.945, \qquad R^2=0.106,\\
    A&=&608.281 R_g + 2042.065, \qquad R^2=0.541,\\
    \Nring&=&25.862 R_g - 169.784, \qquad R^2=0.059.
\end{eqnarray}
\vspace{0.5\baselineskip}
\end{minipage}

\begin{figure} [!htbp]
    \centering
    \includegraphics[width=0.7\linewidth]{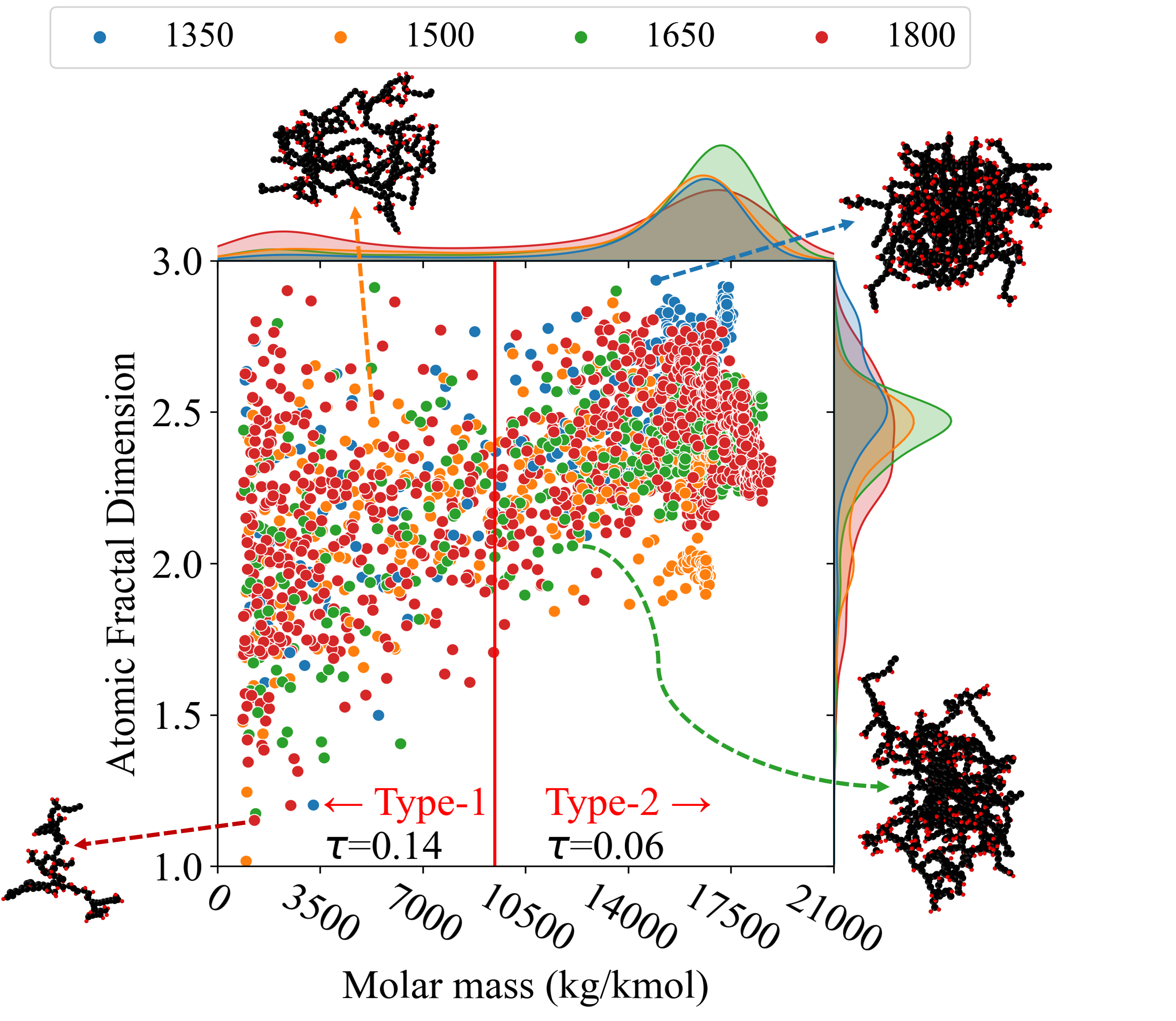}
    \caption{Evolution of atomic fractal dimension ($D_f$) with molar mass} \label{f:DF}
\end{figure}

Fig. \ref{f:DF} shows the comparison between the molar mass and the atomic
fractal dimension ($D_f$) in the form of a joint scatter plot. As discussed
 in Eq. \ref{e:df}, the atomic fractal dimension is calculated using box
counting method for individual incipient particles, not for an aggregate. The
value of atomic fractal dimension ($D_f$) is an indication of the shape of the
incipient primary particles. A value of $D_f$ close to 3 indicates a spherical
shape, while a value of $D_f$ close to 1 indicates a linear shape. Contemporary
engineering-scale soot models often assume spherical incipient particles
\cite{Frenklach1991Jan,Frenklach2002May,Wang2016Feb}.

For both type 1 and type 2 particles, we do not observe any specific
correlation between molar mass and atomic fractal dimension as indicated by the low
$\tau$ value in Fig. \ref{f:DF}.  As the molar mass increases, the atomic fractal dimension
increases and approaches 3. This is because as the incipient particles grow,
the shape of the particles becomes more spherical.
The value of $D_f$ captures the evolution of the shape of a single incipient
particle before coalescence. In the inset of Fig.~\ref{f:DF}, some representative incipient
particles are shown at various stages of their evolution.

\begin{figure}[!htbp]
	\centering
	\subfigure[$R_g$ vs $R_{eq}$]{%
		\resizebox*{0.32\linewidth}{!}{\includegraphics{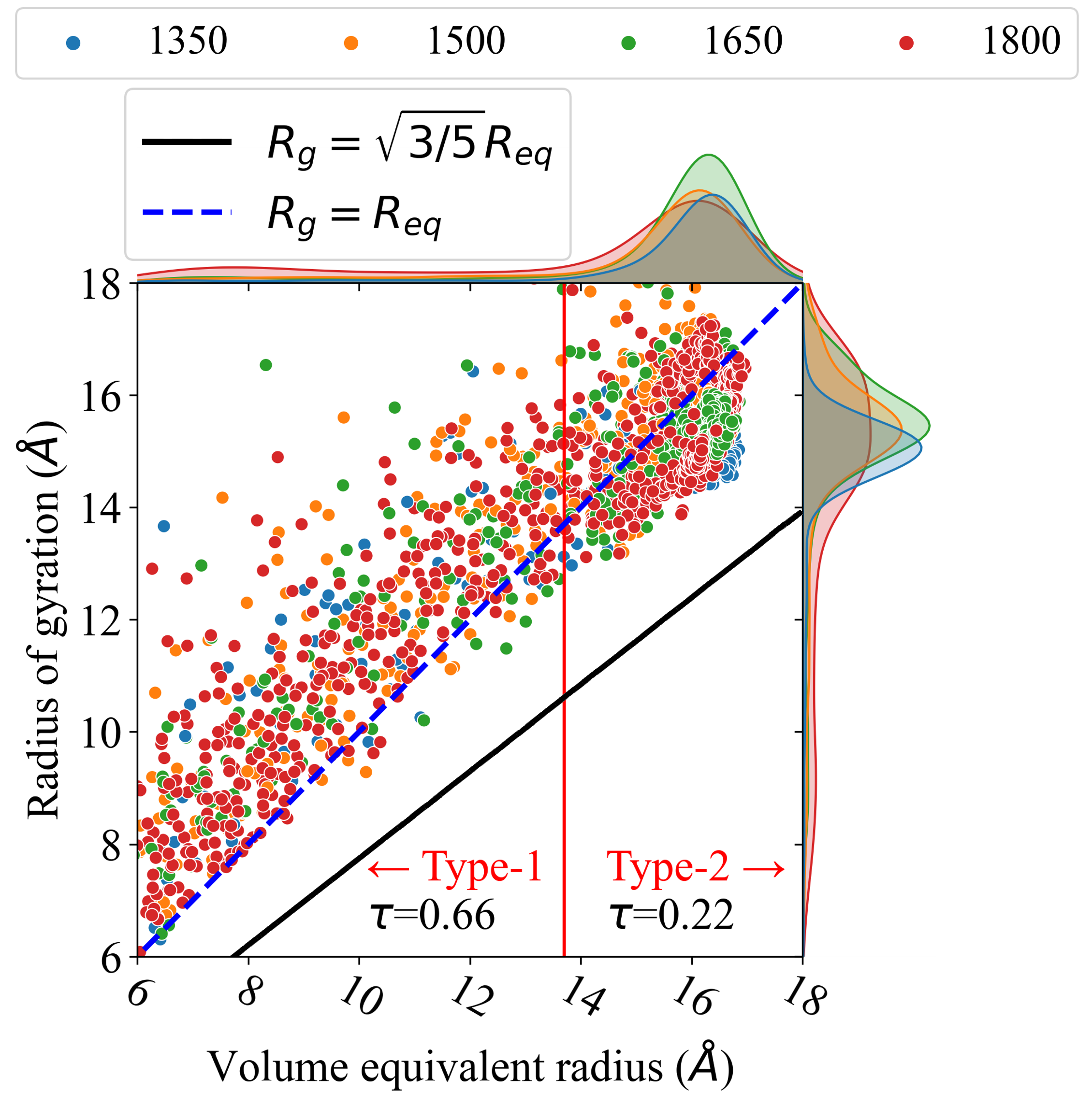}\label{f:Rg_Re}}}
	\subfigure[$\frac{R_g}{R_{eq}}$ vs $N_c$ for type 1]{%
		\resizebox*{0.32\linewidth}{!}{\includegraphics{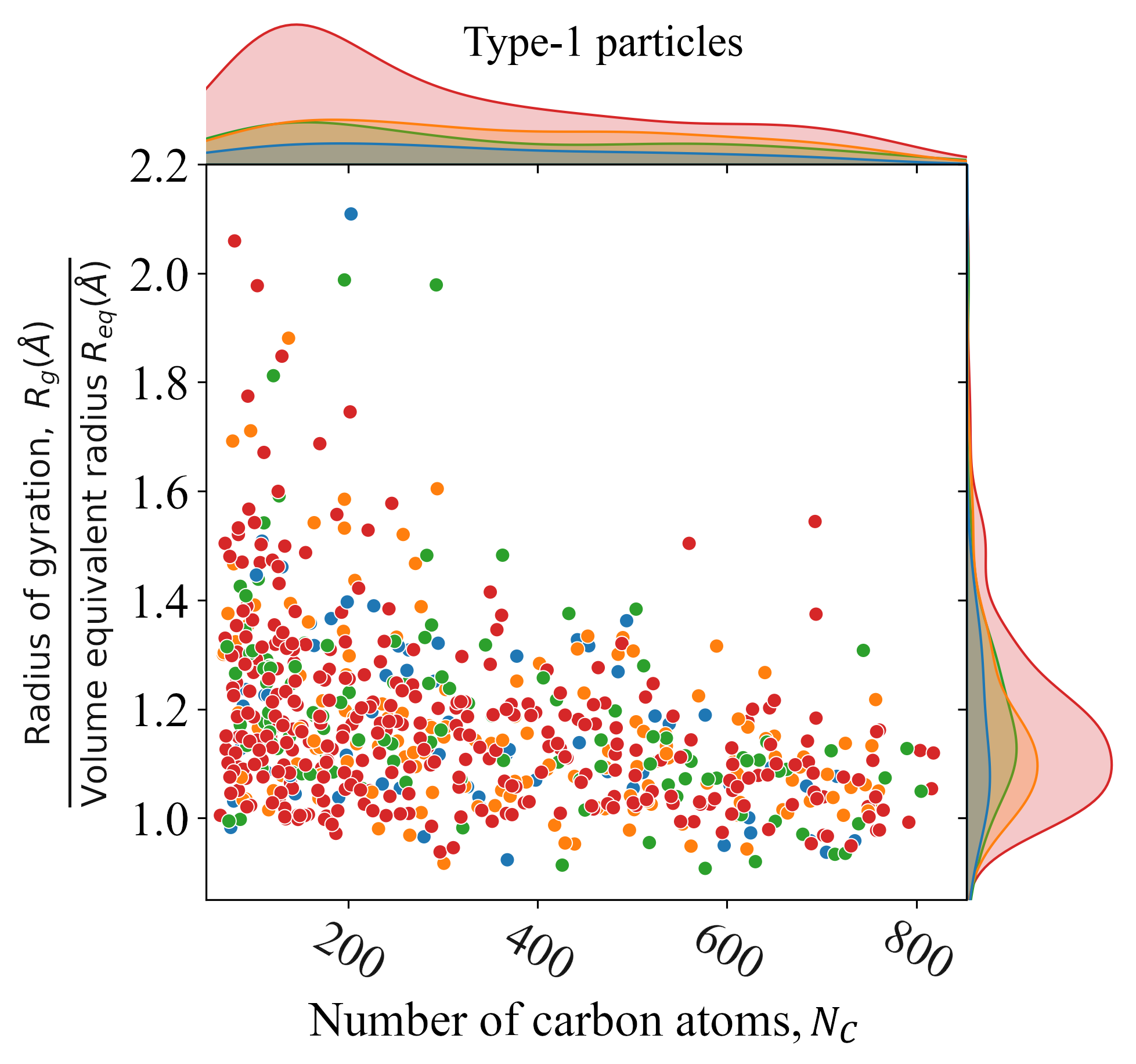}\label{f:t1RGEQ}}}
  \hfill
	\subfigure[$\frac{R_g}{R_{eq}}$ vs $N_c$ for type 2]{%
		\resizebox*{0.32\linewidth}{!}{\includegraphics{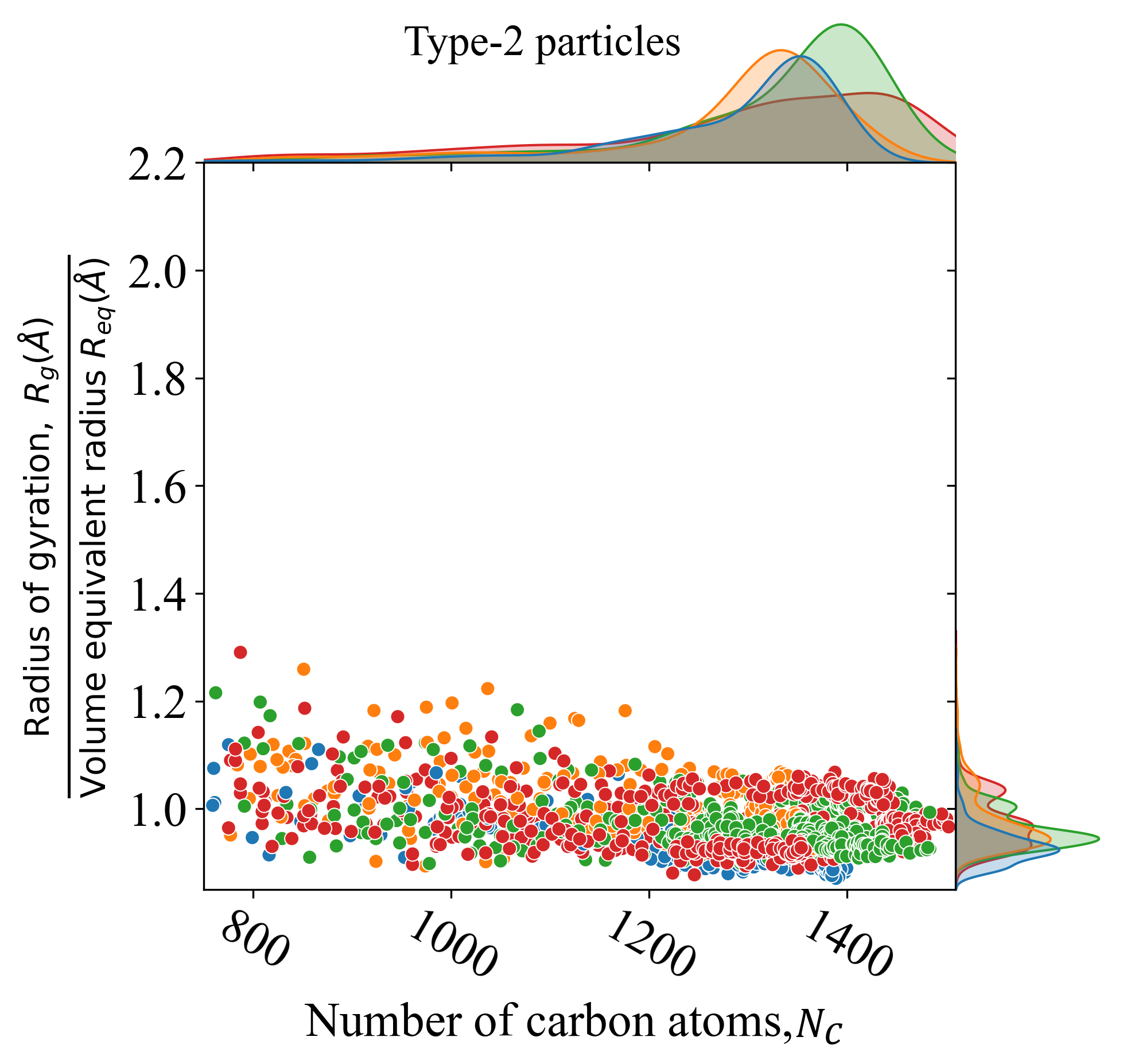}\label{f:t2RGEQ}}}
	\caption{Evolution of radius of gyration ($R_g$) and volume equivalent radius  ($R_{eq}$): Comparison between volume equivalent radius and radius of gyration for all particles (a); ratio of radius of gyration and volume equivalent radius vs number of carbon atoms for type 1 particles (b) and for type 2 particles (c).} \label{f:RGRE_rel}
\end{figure}

The correlation between the volume equivalent radius and the radius of gyration
is shown in Fig. \ref{f:Rg_Re}. The transition from type 1 to type 2
particles occurs around a volume equivalent radius of 13.8~\AA~corresponding to the red
vertical line in Fig.~\ref{f:Rg_Re}. An overall good correlation is
observed between the two quantities for type 1 particles. Like the other
morphological features, the type 2 particles show a comparatively weaker
correlation compared to the type 1 particles. Since the particles tend to become spherical as they evolve (Fig. \ref{f:DF}),
the volume equivalent radius can be a good approximation for the radius of gyration. It should be noted here that although the particles become more spherical, they are not solid spheres, therefore relationships between the  volume-equivalent radius and radius of gyration do not follow the classical \editsf{$\sqrt{\nicefrac{3}{5}}$} scaling (black solid line). Rather \editsf{they} closely follow the $R_g=R_{eq}$ scaling (blue dashed line) as shown in Fig. \ref{f:Rg_Re}.

This becomes more evident if we look at the evolution of the ratio of radius of gyration and volume-equivalent radius ($\nicefrac{R_g}{R_{eq}}$) as the particles grow in size. Figure~\ref{f:t1RGEQ} and \ref{f:t2RGEQ} depict $\nicefrac{R_g}{R_{eq}}$ vs. the number of carbon atoms ($N_c$) for type 1 and type 2 particles respectively. For small type 1 particles, the ratio $\nicefrac{R_g}{R_{eq}}$ shows a wide range of values ranging from 1.0 to 2.1. But within the increase in particle size, i.e., increase in the number of carbon atoms ($N_c$) in the particle, the spread of the  ratio $\nicefrac{R_g}{R_{eq}}$ becomes narrower and approaches unity as seen from Fig. \ref{f:t1RGEQ}. This is because the particles become more spherical and compact as they grow in size. For type 2 particles, the ratio $\nicefrac{R_g}{R_{eq}}$ steadily hovers near 1.0 with very little spread in the data as observed from Fig. \ref{f:t2RGEQ}. This is because the type 2 particles are more graphitized and have a more compact structure compared to type 1 particles.

\subsection{Evolution of ring structures in incipient soot particles} \label{ss:chem_evolution}

\begin{figure}[!htbp]
    \centering
    \subfigure[Number of 5-/6-/7-membered cyclic structures]{%
        \resizebox*{\linewidth}{!}{\includegraphics{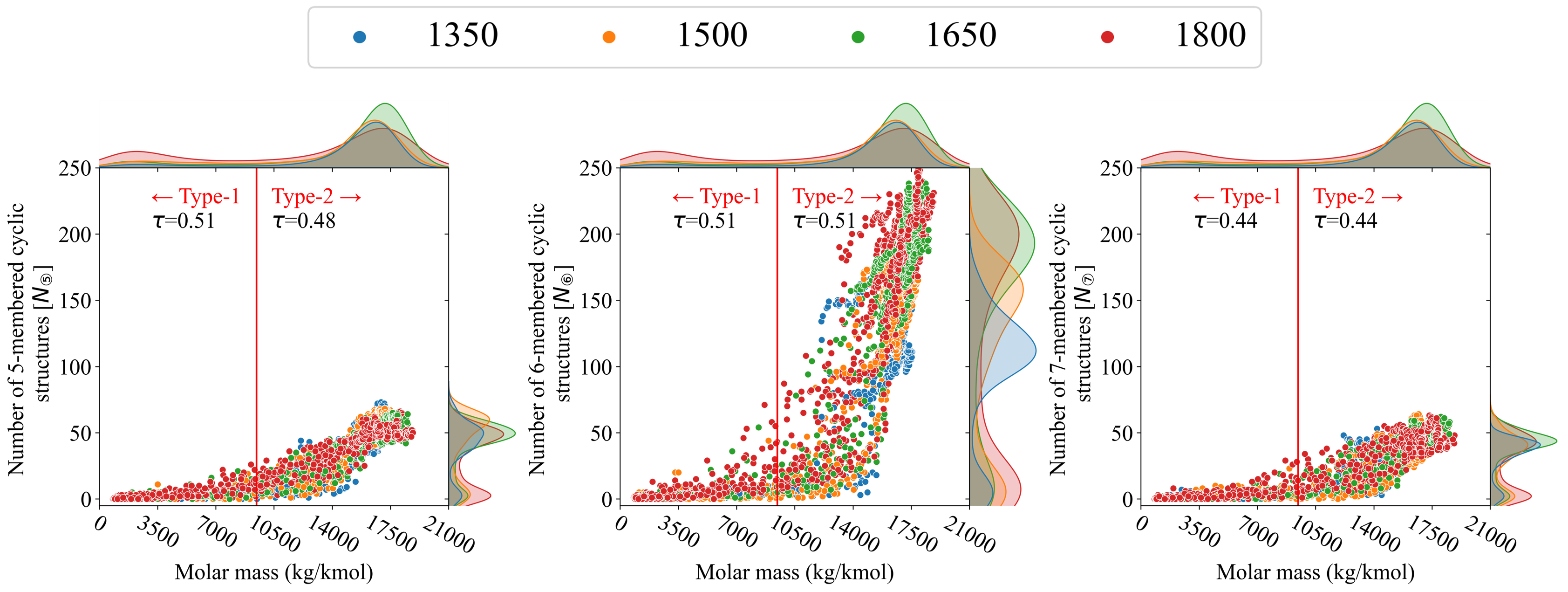}\label{f:ring_total}}}
    \subfigure[Fraction of 5-/6-/7-membered cyclic structures]{%
        \resizebox*{\linewidth}{!}{\includegraphics{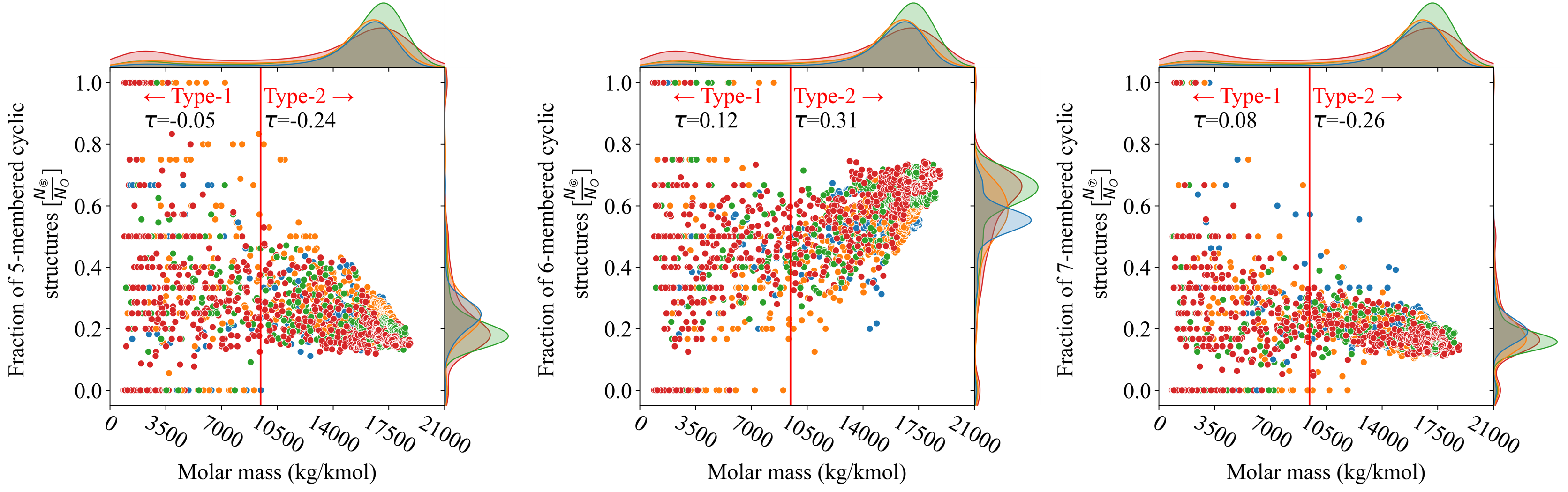}\label{f:ring_frac}}}
    \caption{Evolution of cyclic structures in incipient soot particles.}
    \label{f:rings}
\end{figure}

The ring structures in soot particle play an important role in their
evolution and impact their physical and chemical properties~\cite{Salamanca2020Sep}. As seen in
Figs.~\ref{f:Mass_Rel} and \ref{f:rg_Rel}, the number of rings in  incipient soot particles evolve differently from the morphological features. Therefore, we looked into the evolution of
5-/6-/7-membered rings more closely in this section. Fig. \ref{f:rings} shows
the evolution of the total number of 5-/ 6-/ 7-membered rings ($\Nfive$,
$\Nsix$, $\Nseven$) in the top row and their fraction with respect to the total number of rings
($\nicefrac{\Nfive}{\Nring}$, $\nicefrac{\Nsix}{\Nring}$,
$\nicefrac{\Nseven}{\Nring}$) in the bottom row as a function of the molar mass of the incipient particles.  The red vertical line indicates the threshold between type 1 and type 2 particles. The type 1 particles are on the left of the line and type 2 particles are on the right. The Kendall's correlation coefficient ($\tau$) between the ring structures with particle mass are reported for each type on their respective zones.

The total number of 5-/6-/7-membered rings in type 1 particles increases \editsf{only slightly} with molar mass and no distinct difference is observed between
different ring structures. However, in the type 2 regime, a rapid increase in
the total number of rings is observed with mass. This increase is most
prominent for the 6-membered rings. The 5- and 7-membered rings also increase
with mass but at a much slower rate.  A higher number of 6-membered rings is observed
with increasing temperature in the type 2 regime as seen from the temperature
markers and density plots on the right side of top middle plot of
Fig~\ref{f:ring_total}.

Figure \ref{f:ring_frac} shows the fraction of
5-/6-/7-membered rings in individual incipient particles and provides insights into how different rings interact and change during the evolution of incipient soot particles. In the type 1 regime of Fig.~\ref{f:rings}, the fraction of 5-/ 6-/ 7-membered rings does not show any order or correlation. Throughout the simulation domain, 5-, 6-, and 7-membered rings form independently due to the chemical interactions between small aliphatics without the influence of other cyclic structures. Once the particles grow enough to transition to type 2, the fractions of 5-/ 6-/ 7-membered rings start to change in an orderly manner. The fractions of 5- and 7-membered rings decrease with mass while the fraction of 6-membered rings increases. This is because the 6-membered rings are the most stable ring structures and as the particle grows, the 5- and 7-membered rings are more likely to break and form 6-membered rings. The fraction of rings is also affected by temperature. A higher fraction of 6-membered rings is observed with increasing temperature in the type 2 regime as seen from the temperature markers and density plots on the right side of the  center plot of Fig.~\ref{f:ring_frac}. The fraction of 5- and 7-membered cyclic structures, on the other hand, decreases slightly with \editsf{increasing} temperature.

\subsection{Summary statistics}\label{ss:summary}
The summary of the statistics is presented in Table~\ref{tab:varSpan} in terms
of mean, standard error to the mean (\texttt{SEM}) and standard deviation (\texttt{SD}) of the extracted features. These values can be useful in determining model parameters and model constants for engineering-scale soot model.
The statistics are
also separated by the type of the particles. It can be seen
that the variation in physical features such as \editsf{simulated} density, radius of
gyration, volume, and surface area is much smaller in type 2 particles than in type 1
particles. 
This indicates that in type 1 regime, particles go through a strong physical and morphological evolution, but in type 2 regime, they go through a chemical restructuring.
Due to these changes, from type to type 2, the atomic fractal dimension increases from 2.1 to 2.5 (more spherical). The proportion of 6-membered rings increases significantly while the fractions of 5-/7-membered rings
decrease as they restructured into the more stable 6-membered rings.

The calculated statistics provide a great validation for the RMD when compared with experimental data reported in the literature.  
Schulz et. al \cite{Schulz2019Jan} reported that the average C/H ratio of soot particles from a premixed ethylene flame was  $2.33\pm0.16$, which matches very well with the mean C/H ratio of \editsf{2.36  with a standard deviation of 0.37 for} the simulated particles in our study. \editsf{It should be noted that, we do see a reduction in the C/H ratio as the particles transition from type 1 to type 2 particles. This can be explained by the internal structure of the incipient particles where the type 2 particles have a larger shell region consisting of mostly non-cyclic aliphatic chains compared to type 1 particles \cite{Mukut_paper1}. }
The average \editsf{simulated} density of the soot particle in our study is in the range of 1.50 (type 1) -- 1.53 g/cm$^3$ (type 2). This is very similar to values reported or assumed in several other studies~\cite{Zhao2007Jan, Veshkini2015Nov, desgroux:hal-03326326}. Furthermore, Johansson et al.~\cite{Johansson2017Dec} \editsf{and \cite{Camacho2015Oct}} found a empirical density of \editsf{about 1.5~g/cm$^3$} for incipient soot particles in their experiments. For PAHs with C/H ratio of 1.0-2.4, Minutolo et. al.~\cite{Minutolo2022Oct} found the average density of soot to be 1.5 gm/cm$^3$ using relationships from~\cite{Michelsen2021Jan}. 

\begin{table}[]
	\centering
	\caption{Mean, Standard Error of the Mean (SEM), and Standard Deviation (SD) of the physicochemical features of soot clusters.}
	\label{tab:varSpan}
	\resizebox{\textwidth}{!}{%
		\begin{tabular}{|p{0.5\textwidth}|c|ccc|c|ccc|c|}
			\hline
            \multirow{2}{*}{\textbf{Property}} &             \multirow{2}{*}{\textbf{Unit}} & \multicolumn{4}{c|}{\textbf{Type 1}} & \multicolumn{4}{c|}{\textbf{Type 2}} \\
            \cline{3-10}
             & & Mean &$\pm$ & SEM & SD & Mean &$\pm$ & SEM & SD\\
             \hline
            \textbf{Simulated Density}                                                                  & $\mathrm{g/cm^3}$                   & 1.502         & $\mathrm{\pm}$          & \multicolumn{1}{c|}{0.005}                   & 0.139                 & 1.533         & $\mathrm{\pm}$          & \multicolumn{1}{c|}{0.001}                  & 0.060                 \\ \hline
			
            \textbf{Radius of gyration}                                                              & \AA                                 & 11.349        & $\mathrm{\pm}$          & \multicolumn{1}{c|}{0.094}                   & 2.430                 & 15.472        & $\mathrm{\pm}$          & \multicolumn{1}{c|}{0.014}                  & 0.741                 \\ \hline
			\textbf{Volume equivalent radius}                                                        & \AA                                 & 9.793         & $\mathrm{\pm}$          & \multicolumn{1}{c|}{0.090}                   & 2.318                 & 16.063        & $\mathrm{\pm}$          & \multicolumn{1}{c|}{0.012}                  & 0.597                 \\ \hline
   			\textbf{Ratio of radius of gyration over volume equivalent radius} &     --                                & 1.176         & $\mathrm{\pm}$          & \multicolumn{1}{c|}{0.007}                   & 0.171                 & 0.964         & $\mathrm{\pm}$          & \multicolumn{1}{c|}{0.001}                  & 0.05                  \\ \hline
			\textbf{Molar mass}                                                                      & kg/kmol                             & 4073.8        & $\mathrm{\pm}$          & \multicolumn{1}{c|}{99.9}                    & 2584.0                & 16104.5       & $\mathrm{\pm}$          & \multicolumn{1}{c|}{36.2}                   & 1866.3                \\ \hline
			\textbf{Volume}                                                                          & $\mathrm{\AA^3}$                    & 4600.3        & $\mathrm{\pm}$          & \multicolumn{1}{c|}{116.3}                   & 3008.3                & 17432.3       & $\mathrm{\pm}$          & \multicolumn{1}{c|}{35.6}                   & 1834.6                \\ \hline
			\textbf{Surface area}                                                                    & $\mathrm{\AA^2}$                    & 3090.6        & $\mathrm{\pm}$          & \multicolumn{1}{c|}{64.0}                    & 1654.9                & 7369.2        & $\mathrm{\pm}$          & \multicolumn{1}{c|}{11.9}                   & 612.2                 \\ \hline
			\textbf{Atomic fractal dimension}                                                        &                                     & 2.093         & $\mathrm{\pm}$          & \multicolumn{1}{c|}{0.013}                   & 0.340                 & 2.460         & $\mathrm{\pm}$          & \multicolumn{1}{c|}{0.003}                  & 0.169                 \\ \cline{1-1} \cline{3-10}
			\textbf{Fraction of 5-membered rings}                                                    &                                     & 0.374         & $\mathrm{\pm}$          & \multicolumn{1}{c|}{0.011}                   & 0.294                 & 0.218         & $\mathrm{\pm}$          & \multicolumn{1}{c|}{0.001}                  & 0.063                 \\ \cline{1-1} \cline{3-10}
			\textbf{Fraction of 6-membered rings}                                                    & --                            & 0.399         & $\mathrm{\pm}$          & \multicolumn{1}{c|}{0.010}                   & 0.258                 & 0.605         & $\mathrm{\pm}$          & \multicolumn{1}{c|}{0.002}                  & 0.084                 \\ \cline{1-1} \cline{3-10}
			\textbf{Fraction of 7-membered rings}                                                    &                                     & 0.227         & $\mathrm{\pm}$          & \multicolumn{1}{c|}{0.009}                   & 0.221                 & 0.177         & $\mathrm{\pm}$          & \multicolumn{1}{c|}{0.001}                  & 0.041                 \\ \cline{1-1} \cline{3-10}
		    \hline
        \end{tabular}%
	}
\end{table}

\begin{figure} [!htbp]
    \centering
    \includegraphics[width=0.8\linewidth]{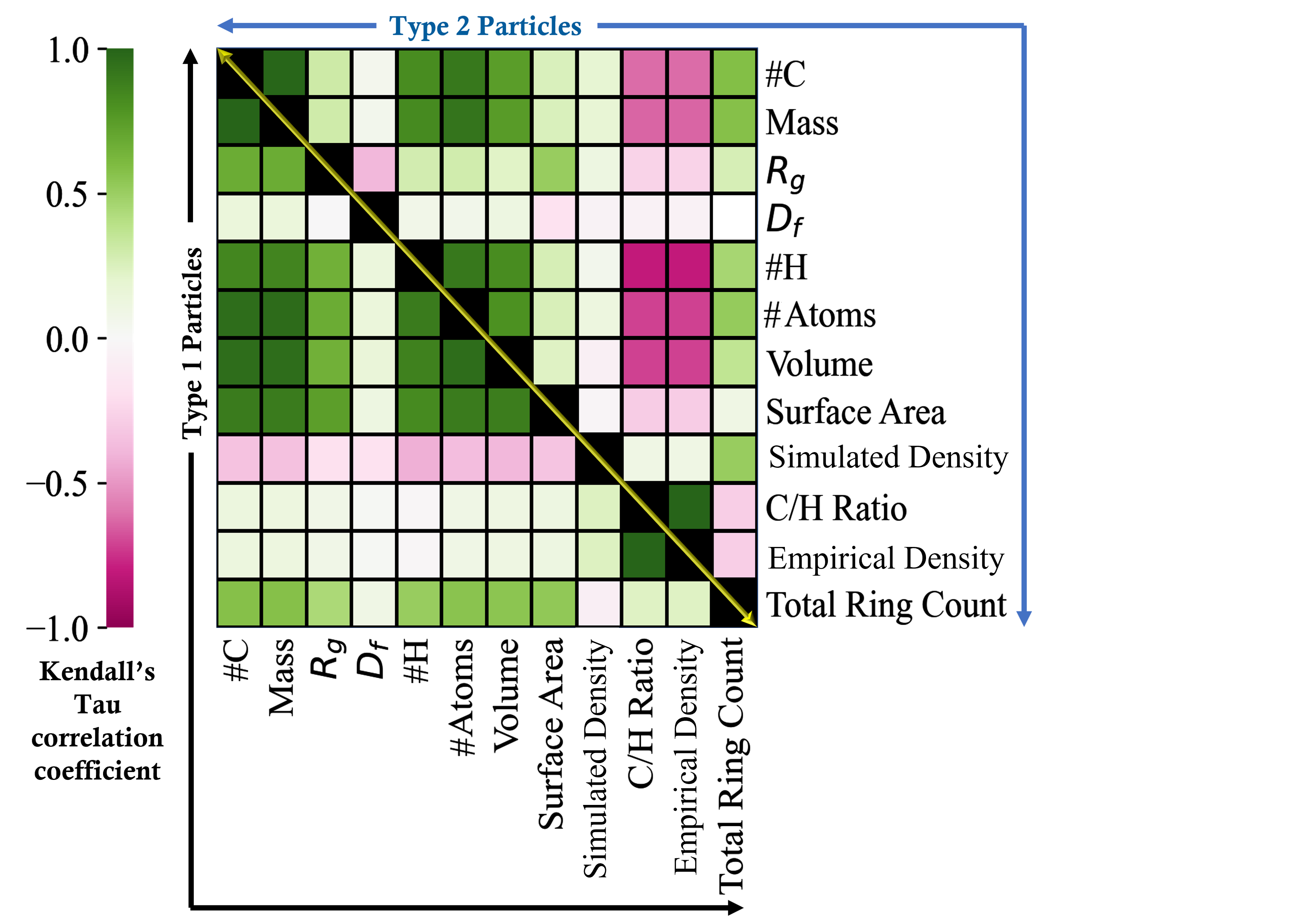}
    \caption{Kendall's rank correlation coefficient between different physicochemical features of type 1 (lower-left triangular matrix) and type 2 (upper-right triangular matrix) incipient soot particles.}
    \label{f:heatmap}
\end{figure}

Finally, a summary correlation matrix between each pair of features for both type 1 and type 2 incipient particles are presented in Fig.~\ref{f:heatmap}. The correlation matrix is a square matrix with the number of rows and columns equal to the number of properties. The diagonal elements of the matrix are always 1,
and the off-diagonal elements are the correlation coefficients between the corresponding properties. The correlation coefficients are calculated using Kendall's Tau test. The correlation coefficients are presented in the form of a heat map in Fig. \ref{f:heatmap}, where the color intensity is proportional to the value of the correlation coefficient. Both type 1 and type 2 correlations are presented in the square correlation matrix. The lower triangular region represents type 1 particles and the upper triangular region represents type 2 particles. For simplicity, the correlation matrix is presented only for a subset of features explored.

If the nature of correlations were to be the same between the types of particles, the
lower and upper triangular portions would be symmetric about the diagonal.
However, as seen in Fig~\ref{f:heatmap}, the quality and trends of correlations
change significantly between the types of particles. For example, the mass
and empirical density is negatively correlated in type 2 particles
whereas they are almost uncorrelated in type 1 particles. A correlation matrix like
this can be useful in creating a reduced-order soot model in the future.

\section{Conclusion} \label{conclusion}

A series of molecular dynamics simulations was performed to study the evolution
of incipient soot particles during acetylene pyrolysis at four different
temperatures. The evolution of the incipient soot particles was studied via
their physicochemical characteristics such as mass, volume, surface area,
radius of gyration, density, and the number of 5-/6-/7-membered rings. Using
unsupervised machine learning techniques, the incipient soot particles are
classified into two types -- type 1 and type 2 -- based on their morphological
and chemical features. This classification was very well predicted
by the number of carbon atoms in the particles, indicating the two types corresponding to early and late stages of incipient soot. \editsf{The RMD-derived incipient soot particle density and C/H ratio shows excellent agreement with experimental data. }

The following conclusions were drawn from the study.
\begin{enumerate}
    
    \item Morphological features of incipient particles are often well correlated with
          each other and with the size of the particle (indicated by the molar mass, number of
          atoms, or radius of gyration). However, the quality and nature of their
          correlations are different between different types of particles.
    \item Type 1 particles usually show stronger correlations between different
          morphological features and size than type 2 particles. Morphological features in type 2 particles show much smaller variation than type 1 particles.
    \item Morphological features of incipient particles, \editsf{e.g., volume, surface area, radius of gyration etc.} are not affected by
          temperature. However, the chemical characteristics, \editsf{e.g., number of cyclic structure and 5-,6-,7-membered rings} of incipient particles are affected
          by temperature.
    \item At the early stage of incipient soot (type 1), 5-, 6-, and 7-membered rings are
          formed independently from one another. However, at the later stage (type 2), the 5- and 7-membered rings are more likely to break and form 6-membered rings.
    \item The incipient particles evolve into a more spherical shape as they transition from type 1 to type 2.
    \item Overall, type 1 particles show a more prominent morphological evolution \editsf{(i.e. volume, surface area and radius of gyration)} than type 2 particles \editsf{with increasing particle mass}. On the other hand, type 2 particles go through a more prominent chemical evolution \editsf{(i.e. evolution of 5-,6-, and 7-memebered cyclic structures)} than type 1 particles. 
\end{enumerate}

The estimation of various quantitative properties and their correlation in incipient soot reported in this work can improve engineering-scale models for soot inception and growth.

\section{Acknowledgments}
The research benefited from computational resources provided through the NCMAS, supported by the Australian Government, The University of Melbourne’s Research Computing Services and the Petascale Campus Initiative.  K.M.M. and S.P.R. acknowledge funding support from the National Science Foundation as some of this material is based upon work supported by the National Science Foundation under Grant No. 2144290. 

\bibliographystyle{elsarticle-num}
\bibliography{main}
\end{document}